%% file: 000-Main-File.tex
\documentclass[sigconf]{acmart}

\usepackage{multirow}
\usepackage{colortbl}

\graphicspath{{figures/}}
\newcommand{\imgw}{\linewidth}

\usepackage{soul}
\usepackage{url}
\usepackage{hyperref}

\AtBeginDocument{%
  }

\copyrightyear{2026}
\acmYear{2026}
\setcopyright{cc}
\setcctype{by}
\acmConference[ASSETS '26]{The 28th International ACM SIGACCESS Conference on Computers and Accessibility}{October 25--28, 2026}{Vila Nova de Gaia, Portugal}
\acmBooktitle{The 28th International ACM SIGACCESS Conference on Computers and Accessibility (ASSETS '26), October 25--28, 2026, Vila Nova de Gaia, Portugal}
\acmDOI{10.1145/3797867.3829036}
\acmISBN{979-8-4007-2521-0/2026/10}

\begin{document}


\title[Q\&A or Document-Based for Screen Reader Users?]{Q\&A or Document-Based? The Effects of Interface Type on How Screen Reader Users Access Interconnected Documents}


\author{Colleen F. Cipriano}
\affiliation{%
  \department{Department of Computer Science}
  \institution{University of Victoria}
  \city{Victoria, British Columbia}
  \country{Canada}}
\email{colleencipriano@uvic.ca}
\orcid{1234-5678-9012}

\author{Yichun Zhao}
\affiliation{%
  \department{Department of Computer Science}
  \institution{University of Victoria}
  \city{Victoria, British Columbia}
  \country{Canada}}
\email{yichunzhao@uvic.ca}
\orcid{1234-5678-9012}

\author{Miguel A. Nacenta}
\affiliation{%
  \department{Department of Computer Science}
  \institution{University of Victoria}
  \city{Victoria, British Columbia}
  \country{Canada}}
\email{nacenta@uvic.ca}
\orcid{1234-5678-9012}

\author{Kotaro Hara}
\affiliation{%
  \department{School of Computing and Information Systems}
  \institution{Singapore Management University}
  \city{Singapore}
  \country{Singapore}}
\email{kotarohara@smu.edu.sg}
\orcid{1234-5678-9012}

\author{Jaylee Soh}
\affiliation{%
  \department{Lee Kong Chian School of Business}
  \institution{Singapore Management University}
  \city{Singapore}
  \country{Singapore}}
\email{jaylee.soh.2022@business.smu.edu.sg}
\orcid{1234-5678-9012}

\renewcommand{\shortauthors}{Trovato et al.}

\begin{abstract}
  \input{sections/005-Abstract}

\end{abstract}

\ccsdesc[500]{Human-centered computing~Accessibility}
\ccsdesc[500]{Human-centered computing~Empirical studies in accessibility}
\ccsdesc[300]{Human-centered computing~Accessibility technologies}

\keywords{Document Accessibility, Mental Models, Conversational User Interface, Exploratory Search, Information Seeking, Blind People, Screen-Readers}


\maketitle

\input{sections/010-Intro}
\input{sections/022-RelatedWork}
\input{sections/030-Methods}

\input{sections/040-Results}
\input{sections/050-Discussion}
\input{sections/060-Conclusion}


%
\begin{acks}
We would like to thank the Canadian Council of the Blind for their support in connecting us with our participants. We are grateful to the participants who gave their time, insights, and trust to this work. We also acknowledge the broader network of practitioners, advocates, and community members who provided us support and guidance. 

This research is supported by the University of Victoria, NSERC DG 2020-04401 and the Singapore Ministry of Education (MOE) Academic Research Fund (AcRF) Tier 1 grant (Project ID: 24-SIS-SMU-039).
\end{acks}

\bibliographystyle{ACM-Reference-Format}
\bibliography{000-bibliography,references_m_zotero}











\end{document}

%% file: sections/005-Abstract.tex
Blind and low-vision (BLV) users are increasingly engaging with large language model (LLM) interfaces to access documents,  but it is unclear how such systems support or hinder their ability to build interconnected knowledge. To examine this gap, we compared a Question–Answer Interface (QAI) that supports open-ended conversational inquiry, with a Document Interface (DI) based mostly on traditional structured text document navigation. We recruited 16 BLV screen reader users where they used both interfaces to explore two fictional worlds. Data from interaction logs, concept maps, decision-based tasks, and semi-structured interviews provide comparative insights into how interface design supports knowledge construction. Findings show that participants visited more distinct documents with the DI and formed larger and more correct mental models with the DI than with the QAI. They were also more able to apply the knowledge they had gained. Simultaneously, many still preferred the QAI and often estimated that they had explored more, formed better mental models and applied their models better when acquiring the information with the QAI, despite this not being the case. Our analysis suggests possible interface design reasons for these differences and highlights some of the risks introduced by using question-answer interfaces to access information spaces.

%% file: sections/010-Intro.tex
\section{Introduction}
The conversational approach to browsing information enabled by generative AI (GenAI) has transformed how people can explore document collections. This emergent tool could benefit those who use screen readers (SRs), most often those coming from the Blind and Low Vision (BLV) community. Traditionally, blind users have relied on mainstream browsing technology and assistive technologies like SRs to access information~\cite{asakawa_1998, lunn_2011, perera_sky_2025}. Despite extensive efforts to establish accessibility guidelines and develop automated accessibility checkers, many documents remain inaccessible~\cite{jordan_2024}. Even technically accessible documents can pose challenges when visual information lacks proper context or description, leaving SR users unaware of content on the page~\cite{bigham2017}. Some expect that GenAI tools could eliminate these barriers~\cite{EY2024GenAI}; poor alt text becomes less problematic when AI can describe images directly, and one could find information quickly just by prompting, without having to search through lengthy reports.

However, current AI-powered tools commonly used by BLV users (\textit{e.g.,} ChatGPT~\cite{atchen_2025, adnin_i_2024}) often provide targeted answers to pointed questions. This approach risks creating biased access patterns, systematically directing users toward certain types of content while obscuring others~\cite{adnin_i_2024, sharma_before_2025}. This, in turn, could limit users to partial understanding of complex information in a given collection of documents~\cite{tang_everyday_2025}. That is, we risk creating new accessibility barriers, much like how missing accessibility tags or image descriptions currently hide information for BLV users~\cite{gubbi_mohanbabu_context-aware_2024, doore_images_2024}, the design of information access technology could require additional effort into achieving comprehensive document understanding~\cite{adnin_i_2024}.
Better understanding of how GenAI tools influence BLV people's information exploration and consumption behavior is crucial for the future design of accessible technologies~\cite{tang_everyday_2025, gonzalez_penuela_towards_2025}.

The overarching goal of this paper is to compare traditional document browsing tools with LLM-driven conversational access tools and understand how SR users explore documents, comprehend the information they contain, and assess which aspects of each tool influence preference. 
We call the former a \textit{document-based interface} (DI) and the latter a \textit{question and answering interface} (QAI). We pose three research questions around these interface types.

\textbf{(RQ1)} \textit{How does the difference in interface type affect the patterns by which SR users browse documents?} 

\textbf{(RQ2)} \textit{How does the difference in interface type influence how SR users integrate and apply knowledge from documents?} 

\textbf{(RQ3)} \textit{What aspects of the interface influence SR users' preference over the other?}

To answer these questions, we conducted a study with 16 SR users to examine how they interact with a DI and a QAI when exploring multimodal documents about complex, information spaces. We created two fictional worlds, \textit{Solana} and \textit{Dominion}, with corresponding document sets. Our mixed-methods study comprised of four phases. In Phase 1, we logged participants' document browsing patterns with both interfaces. In Phase 2, we asked participants to engage in concept mapping activities to examine how they construct mental models using different interfaces. In Phase 3, participants applied their mental models by answering a decision-based scenario within the worlds. Lastly, in Phase 4, we conducted interviews to record their overall experience and preferences. 

Our results showed that participants covered a wider range of documents with the DI, whereas the QAI supported more broad connections transitions within a smaller set of documents. Participants produced larger and broader concept maps with fewer errors with the DI, although the QAI concept maps were more densely linked. Interestingly, most participants did not detect these differences, as some preferred DI's explicit structure and QAI for its conversational style. The subjective perceptions of the two interfaces were divided, suggesting divergence in exploration behaviors, strategies, and experiences. 


This study primarily contributes: (i) an empirical examination of how SR users browse and synthesize multimodal information using a DI and a QAI; and (ii) insights on how these information access interface types shape knowledge construction and application.
Given how long SR users have depended on mainstream browsing tools, we seek to advance understanding of how GenAI can influence and change the future of accessible information access. 



%% file: sections/022-RelatedWork.tex
\section{Background and Related Work}
We first introduce the concept of mental model, which we extensively leverage in this paper to conceptualize how people comprehend documents and to operationalize our measurement of this comprehension.
Then we review document-based interfaces and conversational user interfaces, and the efforts in making these interfaces accessible for SR users. Finally, we introduce how GenAI technologies have enabled the conversational user interfaces.

\subsection{Mental Models and Mental Model Elicitation}
We define mental model (based on~\cite{norman-ObservationsMentalModels-1983, gentner-MentalModels-2014, simonFormsMentalRepresentation1978}) as \emph{an internal representation that allows a person to organize knowledge gained about the world and act in consequence to this knowledge}. Mental models are commonly used to examine cognition in applied and theoretical areas (\textit{e.g.,}~\cite{jonesMentalModelsInterdisciplinary2011, saucierMentalModelsApproach2025}). 
A mental model offers a lens through which we can examine cognitive activities that are otherwise hard to observe. For example, learners construct new mental models ``when confronted with new learning tasks''~\cite{bucciarelliMentalModelsImproving2012, nadkarniInstructionalMethodsMental2003a}. 
Methods for extracting these models include concept mapping and card sorting, each appropriate for different scenarios~\cite{harperContextDrivenFrameworkSelecting2019}. 
Prior work on blind users' mental models has established that non-visual interaction is shaped by the structure of the internal representation of a system, but has mostly examined desktop environments and screen readers through verbal protocols, observations, and performance data, rather than externalized representations~\cite{kurniawan_mental_2002, abidin_blind_2012}. Moreover, work in this area emphasizes the difficulty of eliciting mental models directly~\cite{saei_mental_2010}.


In this paper, we use concept mapping as a way to assess the cognitive process of SR users as they engage with a corpus of documents and develop comprehension. Concept mapping is a way to externalize how people organize, connect, and build comprehension by representing concepts as nodes and their relationships as links~\cite{bias_concept_2015, gerken_concept_2011, sanchez_concept_2010}. Adopting on earlier critiques that verbal protocol alone is incomplete~\cite{abidin_blind_2012}, concept mapping allowed us to examine construction of internal representations, beyond just measuring retrieval from memory.



\subsection{Accessing Documents and Information Access Technologies}

In traditional information retrieval literature, users' tasks in accessing a corpus of documents are broadly divided into search and browsing~\cite{baeza-yates_modern_nodate,croft_search_2010,agosti_information_2008, saracevic1997}.
In search, the goal is to find relevant information or a list of documents based on a search query~\cite{preininger_differences_2021,kim_conversations_2025}. 
In its most basic form, search prompts a user to enter a query, and the information access technology (\textit{e.g.,} a search engine) returns the results~\cite{lewis_natural_1996}.
While search has evolved to accommodate more natural language queries to handle greater ambiguity~\cite{moore_natural_2018}, and support iterative refinement conversationally with the use of \textit{conversational user interfaces} (CUIs)~\cite{hall_conversational_2018, kelly_methods_2009, saracevic1997,dhingra_towards_2017, kim_conversations_2025}, the fundamental interaction type of ``query-and-get-results'' paradigm remained essentially unchanged until recently. Browsing represents a more exploratory approach to information access, which is associated with learning and investigative sense-making~\cite{marchionini_2006}---activities that are more aligned with the focus of our work. Users typically lack a specific query in browsing; instead, they may seek to understand general concepts within documents or gain an overview of available information before conducting more targeted searches~\cite{athukorala_2016,marchionini_2006}. Browsing occurs through \textit{document-based interaction}. 
When navigating the Web, for instance, sighted people use browsers through direct manipulation, while SR users additionally rely on assistive technologies~\cite{asakawa_1998, lunn_2011}. 

The emergence of GenAI-based tools follows a broader shift in information retrieval toward treating search as a text-to-text transformation problem~\cite{mo_survey_2025,suri_use_2024,yang_searchchat_2025,holtz_new_2024,preininger_differences_2021,sowa_expert_2025}. Compared with traditional CUIs, these systems can interpret nuanced input, generate human-like responses, and---crucially for document access—--synthesize text across multiple documents and maintain conversational context~\cite{ai_foundations_2025,mpgeetha_conversational_2024}. Furthermore, tools like NotebookLM\footnote{\url{https://notebooklm.google.com}} and Elicit\footnote{\url{https://elicit.com}} enable information exploration within document corpora through iterative question-answering, and prior work shows that people do ask open-ended questions to navigate large collections without explicit queries~\cite{atchen_2025, adnin_i_2024}. However, we still know little about what happens when an LLM becomes the intermediate layer between users and documents, especially in accessibility-centered exploration of interlinked documents.

This gap matters because the same properties that make GenAI tools appealing may also reshape access in limiting ways. Beyond general concerns like GenAI inducing overreliance~\cite{passi_overreliance_2022, bainbridge_ironies_1983},
and difficulty and cognitive burden associated with prompt-based interaction~\cite{tankelevitch_metacognitive_2024, dang_choice_2023,lehmann_functional_2024,subramonyam_bridging_2024}, the dominant ``prompt-and-get-summary'' interaction encourages targeted information extraction and may reduce exposure to a wider range of documents~\cite{pucci_defining_2023, furini_conversational_2020, baez_exploring_2022, kodandaram_enabling_2024}. Prior work has linked this to weaker exploration, comparison, and continuity of access~\cite{torres_accessibility_2019, baez_exploring_2022, fast_iris_2018}. Using the stratified model of information retrieval as an analytical lens, we expect that effort is made toward prompt formulation and summary interpretation at the affective layer~\cite{saracevic1997}, while reducing direct interaction with the document structure at the interface layer.


If we unknowingly design GenAI-powered exploration tools that could bias information access, 
we risk creating new barriers for SR users; just as lack of appropriate document tags limits access to information, the design of information access technology may obscure parts of a collection and require additional effort into achieving comprehensive understanding.
Therefore, in this work, we conduct a study comparing two interfaces---document-based interface and CUI-based question and answering interface powered by GenAI---that manifest different types of interaction.

\subsection{Document Accessibility}
Our goal is to understand the effects of interaction types on how SR users explore and comprehend information in documents.
This requires us to isolate the accessibility of documents and interfaces by making them as accessible as possible.

The development of accessible document-based interfaces and study of information access using these tools has been a significant focus for the HCI and accessibility research communities. 
For example, past work has explored ways to present and extend alt-text or image descriptions to enrich screen reader-friendly interaction with static content~\cite{Jung_Mehta_Kulkarni_Zhao_Kim_2022,Zong_Lee_Lundgard_Jang_Hajas_Satyanarayan_2022, Zhao_TADA, Chart_Reader, chheda-kothary_it_2025, zong_rich_2022}.
Tactile graphics offer another means of access: raised-line representations of visual materials that can be explored through touch~\cite{Mukhiddinov_Kim_2021, gupta_tactile_2017}. 
For tactile graphics there are guidelines that can aid in their creation~\cite{wabinski_guidelines_2022, erp_guidelines_2002, butler_technology_2021}. In our study, we created the document-based interface by learning from past interface designs and following existing guidelines. 
To make our document-based interface accessible, we provided sufficient annotations for screen-reading access and used a tactile overlay on a touchscreen device for graphical access.

Early conversational assistants like Alexa and Siri opened the door to voice-based, hands-free information access for SR users. These assistants were appreciated for quick and simple tasks, which lowered barriers to accessing digital content~\cite{oumard_pardon_2022, pradhan_accessibility_2018, abdolrahmani_siri_2018}. Today's LLM-based interfaces can further facilitate access to documents~\cite{perera_sky_2025}. Rather than navigating documents line-by-line, SR users can query LLMs for descriptions, summaries, or clarifications, enabling them to extract and synthesize information more efficiently. This shift in information access allows SR users to direct their navigation path instead of adhering to rigid document structures~\cite{adnin_i_2024}. At the same time, LLMs may also introduce problems by limiting access to a subset of content, making it difficult to ensure complete rather than fragmented understanding~\cite{pucci_defining_2023, furini_conversational_2020, baez_exploring_2022, kodandaram_enabling_2024}. 

%% file: sections/030-Methods.tex
\section{Study Methodology}
To address the RQs from the Introduction, we designed an empirical study to observe SR users navigating unfamiliar information spaces, building mental models of those spaces, and applying the gained knowledge. The study took place over two laboratory sessions in which SR users worked with two interfaces, one based on traditional hyperlinked documents (the Document Interface---DI), and one based on a question-answer paradigm powered by an LLM---QAI). The laboratory setting allowed us to gather detailed data and control our observations. Participants experienced both interfaces (a within-subjects design), allowing explicit comparison. 




\begin{figure*}[t]
  \centering
  \includegraphics[width=\imgw]{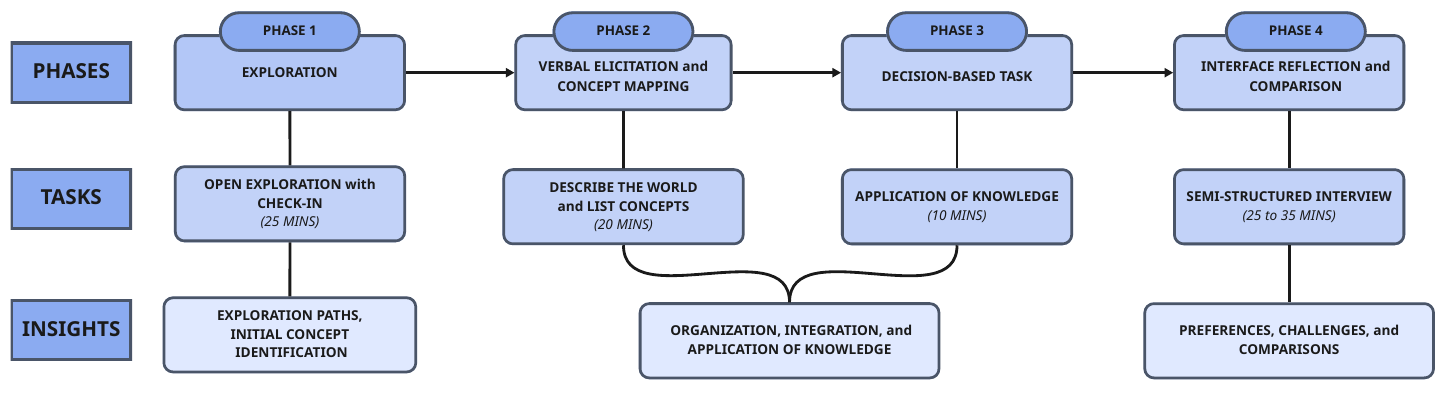} 
  \Description{Horizontal process map of the study. The top row shows four phases: Exploration, Verbal Elicitation and Concept Mapping, Decision-Based Task, and Interface Reflection and Comparison. The middle row lists the activity and duration for each phase. The bottom row summarizes the main outcomes, with curved connectors linking concept listing and knowledge application to the shared insight on organization, integration, and application of knowledge.}
  \caption{Overview of the study procedure and measurements. 
  Participants completed all phases for each of the DI and QAI conditions in a counterbalanced order. At the end of the second session in Phase 4, participants completed an additional set of interview questions comparing the interfaces.}
  \label{fig:study-phases}
\end{figure*}

\begin{figure}[t]
  \centering
  \includegraphics[width=\imgw]{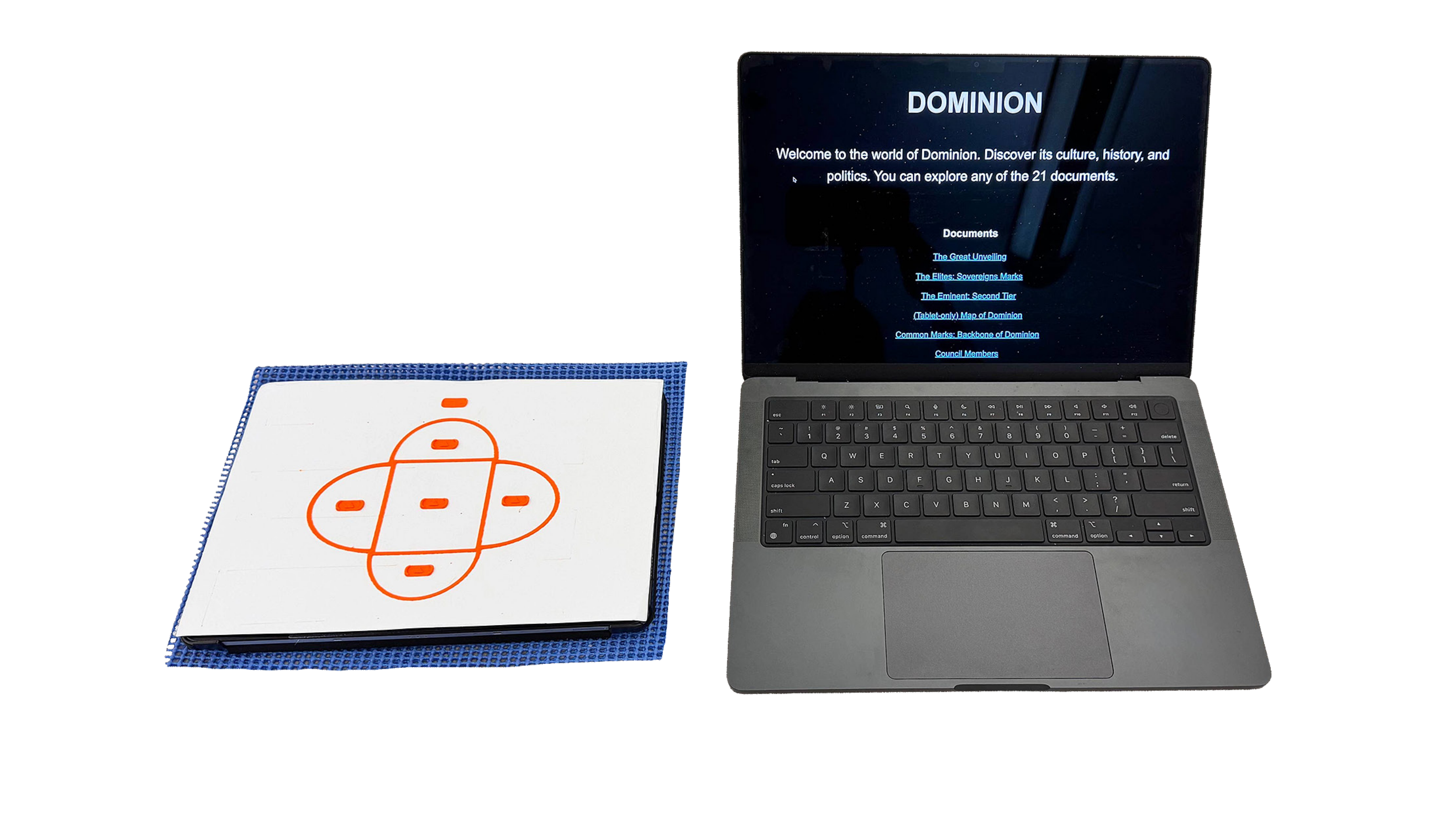} 
  \Description{Setup for the DI condition. On the left, a tablet with a tactile overlay is placed on a non-slip mat next to a laptop, on the right, displaying the Dominion interface.}
  \caption{Experimental setup for the DI condition showing the tactile overlay on top of the tablet (left) on a non-slip mat, with the laptop showing the Dominion interface (right).}
  \label{fig:experimental-setup}
\end{figure}

\subsection{Study Materials: Fictional Worlds}
An important concern in this kind of investigation is the large potential effect of familiarity on people's behavior and understanding. For this reason, we created two information spaces from scratch, guaranteeing no familiarity. These take the form of \textit{interconnected documents}, similar to Wikipedia's hyperlinked pages but describing fictional worlds. The document sets followed the following criteria: (i) minimize the influence of participants' prior knowledge, (ii) encompass sufficient diversity of information and of document types to reproduce real-life complexity, and (iii) provide sufficient depth for extended exploration. We iteratively composed the \textit{Solana} and \textit{Dominion} worlds in a research approach related to Kieras et al.'s~\cite{kieras_role_1984}, informed by existing methodologies in fictional world building~\cite{fischer_building_2021}. 

Each document set included 25 documents, including an index. Out of the 25 documents of each world, 8 (Solana) and 6 (Dominion) were diagrams representing spatial or conceptual aspects of the fictional worlds (\textit{e.g.}, a map of the territory and a diagram of family group configurations). The textual representations resemble wikipedia articles, with headings and meaningful inter-document links. Each document describes an aspect of the world, such as history, governance, and culture. 

We included spatial representations because they are common in real documents and are recognized as important to the BLV community~\cite{elmqvist_visualization_2023,Zhao_TADA}. 
To make spatial documents accessible to our participants, we designed 3D-printed tactile overlays following best practices~\cite{butler_technology_2021, gupta_evaluating_2019}, considering BANA guidelines~\cite{noauthor_guidelines_nodate} and current research on tactile representations~\cite{butler_technology_2021, gupta_evaluating_2019}. We iterated over the designs three times to make them representative of the best accessible spatial representations currently available to the community~\cite{he_using_2025}. Although interactive printable tactile overlays are not currently widespread, they are not very different from printed materials that many in the BLV community are familiar with (e.g., embossed tactile prints) and may soon become practical and affordable~\cite{barros_towards_2023}. 
Excluding spatial information from the DI would have artificially constrained the DI from presenting spatial structure in general.
The implications of this choice are discussed in the Limitations Section below. The documents that form the information spaces of both worlds, including the visual diagrams and their 3D-printed versions are open sourced for use in subsequent research and shared in the Supplementary Materials.

\subsection{Interfaces (Main Condition)}
The two interfaces (DI and QAI, shown in Figures~\ref{fig:DI-interface}~and~\ref{fig:QAI-interface}, provided access to the fictional worlds with the interaction paradigms that we want to compare: direct corpus exploration and LLM-mediated question answering. We designed the conditions to 
isolate these access paradigms so that the observed differences could be attributed to the interaction approach. We strove to implement state-of-the-art interfaces that are the best reasonable alternatives available or soon to be available.



\subsubsection{Document Interface (DI)}
\begin{figure}[t]
  \centering
  \includegraphics[width=\imgw]{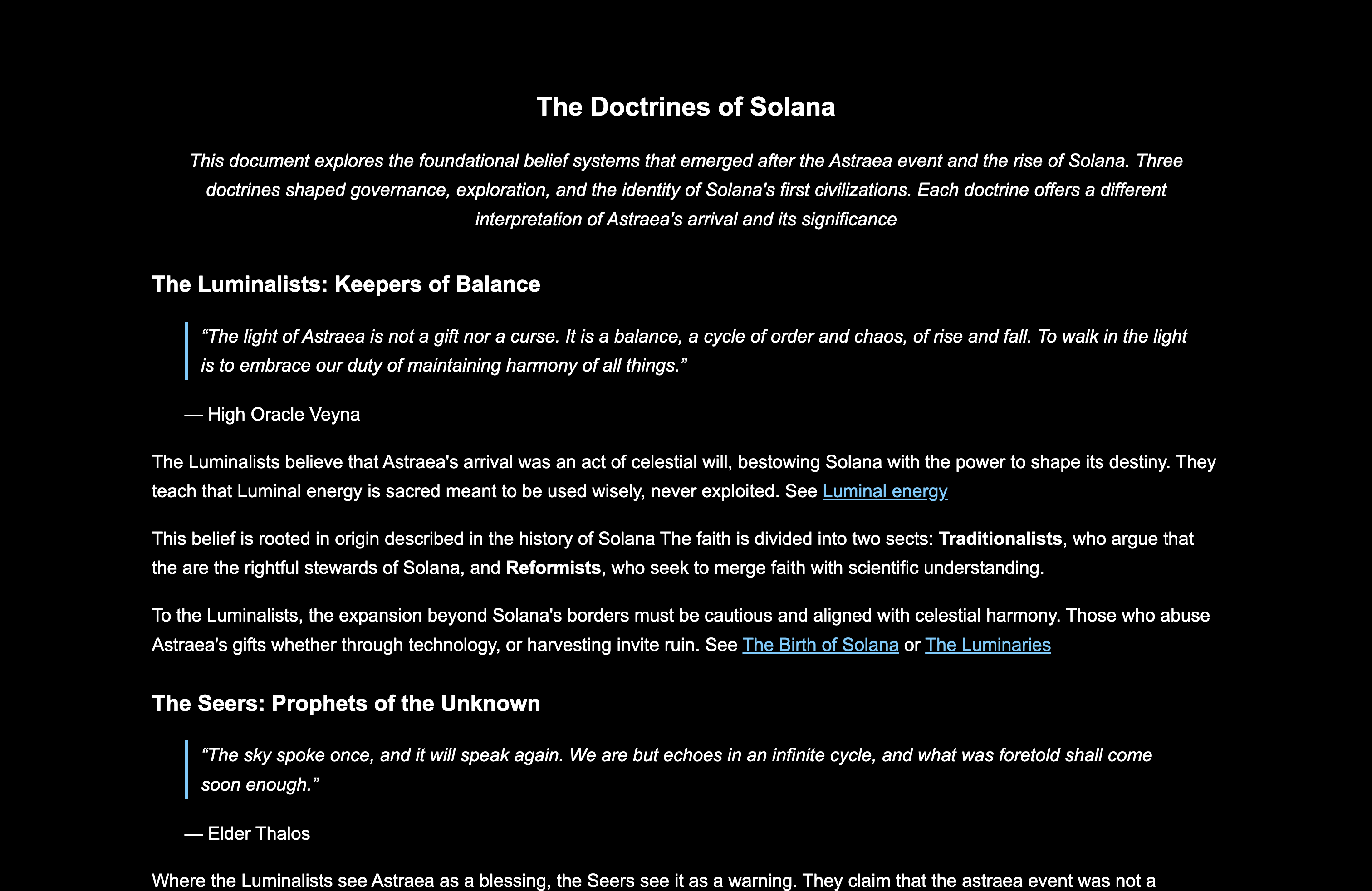} 
  \Description{A sample document in the DI trial. The website page consists of headers, paragraphs, an overview, and a large main content area with a main heading called 'The Doctrines of Solana.' There re also hyperlinks, providing additional references within the document.}
  \caption{An example of one of the documents in the DI interface for the world of Solana.}
  \label{fig:DI-interface}
\end{figure}

\begin{figure}[t]
  \centering
  \includegraphics[width=\imgw]{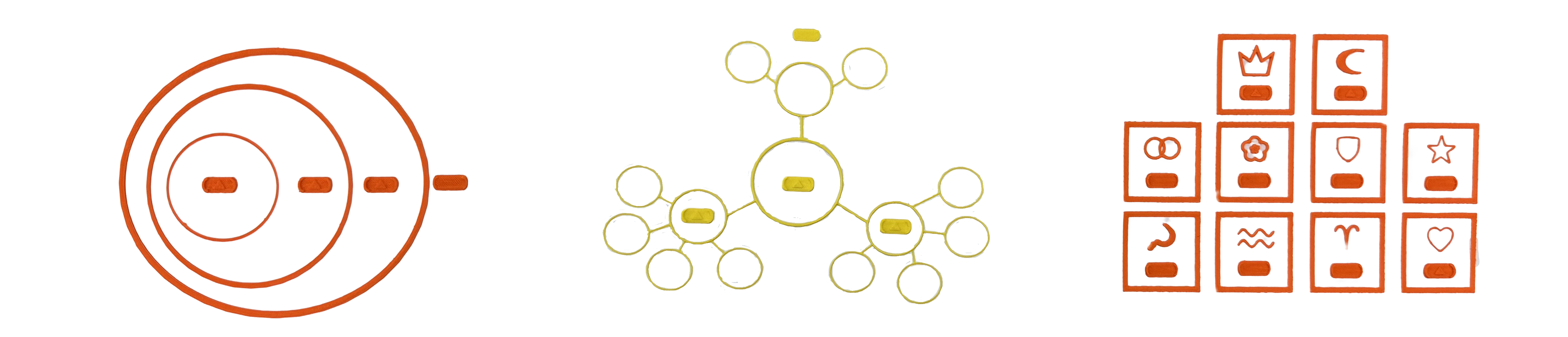} 
  \Description{Tactile overlays. It shows three types of tactile overlays arranged horizontally. The left one represents a map, the center shows a node-link diagram, and the right presents a pyramid of symbols.}
  \caption{Examples of tactile overlays used in the study: a map (left), a diagram of relationships (center), and a set of icons (right).}
  \label{fig:overlays}
\end{figure}

To minimize the need to train participants and maximize ecological validity, participants used their own device and screen reader (SR) to access textual documents. These were web pages meeting the Web Content Accessibility Guidelines---WCAG 2.1~\cite{noauthor_wcag} (see Figure~\ref{fig:DI-interface}). The DI supported index and in-page search with the participants' own SRs.
The spatial documents were accessed as the 3D-printed paper diagrams discussed above, overlaid over a tablet (Samsung Galaxy Tab A9+--- see Figures~\ref{fig:overlays} and~\ref{fig:experimental-setup}). Moving the finger over the overlay produced verbal descriptions of manually labeled areas, in addition to the static relief of the 3D-printed overlay. When a participant followed a hyperlink in a textual document connected to a spatial document, they received a verbal notification to switch to the tablet. If the spatial document had changed, the experimenter manually placed the corresponding document's overlay. The spatial documents also linked to text documents: when tapped, the browser on the main computer jumped to the corresponding document.

The tactile representations preserve spatial relationships that are difficult to communicate through SR navigation alone. Their inclusion reflects our goal of implementing DI as a realistic form of document access. We considered several alternatives for spatial document display such as dynamic pin displays, but decided against because of their low resolution and reliance on Braille, which is not universal among BLV users. Although printable tactile overlays are not currently widespread, they are not very different from printed materials that many in the BLV community are already familiar with (e.g., embossed tactile prints) and may soon become practical and affordable~\cite{barros_towards_2023}.

\subsubsection{Question-Answer Interface (QAI)}
\begin{figure}[t]
  \centering\includegraphics[width=\linewidth,height=.5\textheight,keepaspectratio]{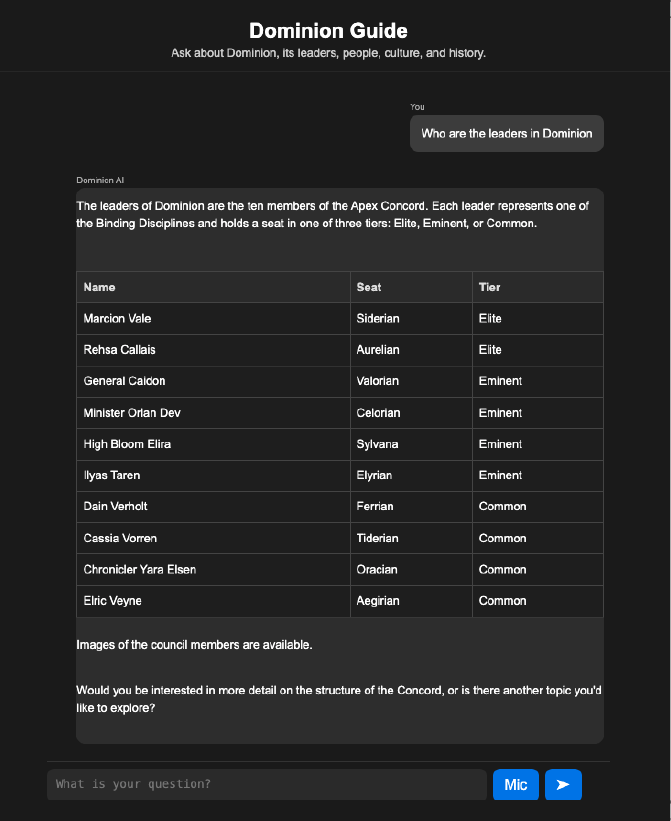} 
  \Description{The QAI interface with the Dominion guide. A user asks, "Who are the leaders in Dominion?" with the system responding with a table listing names, seats, and tiers of ten members. The interface contains a heading and a short description of the Dominion.}
  \caption{The QAI interface for the world of Dominion showing a user query and a structured tabular response.}
  \label{fig:QAI-interface}
\end{figure}

The QAI was a WCAG 2.1-compliant web-based~\cite{noauthor_wcag} chatbot accessible through participants' devices and screen readers.  
Participants could enter questions by typing or using speech input. The system returned information in multiple formats: text, structured bullet points, tables, detailed image descriptions when relevant, and direct quotations from the source documents (Figure~\ref{fig:QAI-interface}). The interface was designed to retrieve and summarize information only from the existing knowledge base.

The underlying conversational system was built using Python with Flask, integrated with Google's Gemini API (Gemini 2.5 Pro). To constrain conversations to information in the documents we used a custom prompt with cache-augmented generation (CAG)~\cite{chan_dont_2025}, loading the corpus of 25 documents to the context window (including the spatial documents). CAG is well-suited to closed knowledge bases because it reduces retrieval latency and 
errors. 
When direct answers were unavailable, the QAI suggested related documents to stay within content boundaries. For out-of-scope questions (\textit{e.g.,} ``Do you have Oreo cookies in Solana?''), it explicitly stated that the information was unavailable, offering relevant alternatives such as ``The archives do not mention that topic. If you're interested, there is detailed information on Solana's economy or its annual festivals.'' We did not observe any instances of hallucination or erroneous responses.

\subsection{Participants}
We recruited 16 participants (age~$\geq$~19) who reported regular use of at least one screen reader (\textit{e.g.,} JAWS, NVDA, VoiceOver, TalkBack). Recruitment took place through community organizations, social media, and snowball sampling. 
Participants reported varied GenAI uses (Table~\ref{tab:demographics}), including accessibility features ($n=11$), work or productivity ($n=8$), technical assistance ($n=7$), writing or content generation ($n=5$), and information seeking ($n=4$). Participants received \$40 per session. The study was approved by the local ethics board. 
\begin{table*}[t]
  \caption{Self-reported demographics of the study participants.}
  \label{tab:demographics}
  \centering
  \small
  \renewcommand{\arraystretch}{1.08}

  \begin{tabular}{
    @{}
    p{0.04\textwidth}
    p{0.07\textwidth}
    p{0.13\textwidth}
    p{0.17\textwidth}
    p{0.30\textwidth}
    p{0.13\textwidth}
    @{}
  }
    \toprule
    \textbf{ID} &
    \textbf{Age} &
    \textbf{Visual ability} &
    \textbf{Occupation} &
    \textbf{GenAI tools used} &
    \textbf{GenAI usage} \\
    \midrule

    \rowcolor{black!5}
    1  & 40--44 & Totally blind & Program coordinator
       & ChatGPT, Copilot, Be My AI & Daily \\

    2  & 19--24 & Totally blind & Voiceover artist
       & ChatGPT, Copilot, Gemini, Be My AI & Daily \\

    \rowcolor{black!5}
    3  & 50--54 & Low vision & Unemployed
       & ChatGPT, Be My AI & Daily \\

    4  & 55--59 & Totally blind & Administrative work
       & ChatGPT, Gemini, Be My AI & 2--3 times a week \\

    \rowcolor{black!5}
    5  & 55--59 & Totally blind & Retired
       & Be My AI & Daily \\

    6  & 25--29 & Totally blind & Unemployed
       & ChatGPT, Be My AI & Daily \\

    \rowcolor{black!5}
    7  & 70--74 & Totally blind & Instructor
       & Be My AI & Daily \\

    8  & 65--69 & Totally blind & Bookkeeper
       & None & N/A \\

    \rowcolor{black!5}
    9  & 55--59 & Low vision & Retired
       & None & N/A \\

    10 & 19--24 & Totally blind & Accessibility tester
       & ChatGPT, Copilot, Gemini, Be My AI & Daily \\

    \rowcolor{black!5}
    11 & 30--34 & Low vision & Marketer
       & None & N/A \\

    12 & 50--54 & Totally blind & Trades
       & ChatGPT, Gemini, Be My AI & Daily \\

    \rowcolor{black!5}
    13 & 65--69 & Totally blind & Administrative work
       & None & N/A \\

    14 & 55--59 & Totally blind & International relations
       & Copilot & Daily \\

    \rowcolor{black!5}
    15 & 55--59 & Totally blind & Client scout
       & Gemini, Copilot & Daily \\

    16 & 45--49 & Totally blind & Administrative work
       & ChatGPT, Gemini, Be My AI & Daily \\

    \bottomrule
  \end{tabular}
\end{table*}

\subsection{Procedure} \label{study procedure}
Each participant completed two 90-minute sessions on separate days at a location of their choice (\textit{e.g.,} public library, participant home, community center). We counterbalanced interface order (DI vs.\ QAI) and world order (Solana vs.\ Dominion) across participants. Given the study duration (Figure~\ref{fig:study-phases}), sessions were separated to 
reduce cognitive load, fatigue, and carryover effects, consistent with prior cognitive work~\cite{keppel_design_1991, hornbaek_whys_2013}. To 
preserve comparative judgments, we capped the interval between sessions at five days~\cite{cepeda_distributed_2006}. In practice, participants completed the second session one to three days later, depending on scheduling and location constraints. 
This also reflects best practices for working with BLV individuals by supporting flexible scheduling, agency in choosing comfortable locations, and use of their preferred device and screen readers. Each session proceeded with four phases in Figure~\ref{fig:study-phases}.


\textbf{Phase 1.} Participants had to follow their interests and openly explore the world using the assigned interface, while developing an overall understanding of the world. We explicitly told them that recall was not the goal and asked them to prioritize acquiring domain knowledge over constrained task performance~\cite{sarrafzadeh_exploring_2014, soufan_towards_2023, marchionini_2006}. Before starting their QAI session, participants were informed that the system has an underlying content of 25 documents. Participants could take notes, but none chose to.

\textbf{Phase 2.} Participants had to describe the world as if to a friend, highlighting what they found most important. From their description, they listed important concepts and explained how concepts relate to each other. Using these responses, the researcher constructed a concept map and asked probing questions to address gaps or overlooked relationships. The map was then validated with the participant and refined as needed to ensure that it accurately reflected their understanding. The researcher followed a structured script and added or connected only concepts and relationships introduced by the participant in order to prevent bias in the analysis. 

Participants did not have access to the assigned interface because doing so would have shifted the evaluation from their mental models to assisted performance. This restricted interface access means that the experimental design aligns with our formulation of RQ2. This phase implements a mental model elicitation technique similar to interview-based concept mapping method~\cite{m_interview_2018}. The assisted concept mapping exercise, in which the experimenter built a graph of concepts and connections is designed to capture mental model organization beyond recall measures~\cite{gerken_concept_2011, bias_concept_2015}. 


\textbf{Phase 3.} Participants applied their understanding by answering a question from a world-specific scenario (\textit{e.g.,} ``As Solana expands, kinship traditions are becoming difficult to maintain. The Luminaries are debating whether to modernize certain laws or preserve Kinship laws strictly. You’ve been asked to propose a policy on this'')
which tested how effectively they could use their gained knowledge. Decision-based scenarios complement the model elicitation procedure of Phase 2 by examining applied reasoning~\cite{paolo_mental_1999, bystrom_task_1995}.  



\textbf{Phase 4.} Participants completed a semi-structured interview about their interactions. At the end of each session, we asked about ease of use, challenges, and overall experience. After the second session, participants additionally answered comparison questions, indicating overall preference, perceived support for exploration and learning, and confidence in answering, accompanied by free-form explanations. 


\subsection{Measurements and Analysis}
The two sessions yielded quantitative and qualitative data to address our research questions. Figure~\ref{fig:study-phases} indicates which measurements correspond to which phases of the experiment.

\subsubsection{Document Traversal Data} \label{traversal data}
We logged the visited documents and transitions across the information space in Phase 1 of each session. 
We analyzed and plotted how participants moved within each interface. 
The quantitative measures extracted from this phase are: 

\begin{itemize}
\item\textit{Unique Document Visits:} A count of distinct documents counted as visited at least once.
\item\textit{Document Visits:} The total number of visits, including revisits.
\item\textit{Unique Document Transitions:} A count of distinct document-to-documenttransitions traversed at least once.
\item\textit{Document Transitions:} The sum of all transitions made between documents.
\item\textit{Transitions per Document:} The transitions normalized by the number of unique document visited.
\end{itemize}

Because the interfaces support different actions, visits and transitions, we necessarily had to operationalize their counts differently across interfaces maintaining equivalence as much possible.
In DI, a \textit{visit} occurs upon opening a document. In QAI, we manually mapped each response to the documents where its information appeared. We counted a source document as visited when the response included information corresponding to five or more contiguous sentences from that document; shorter references were treated as summaries and were not counted as visits. We chose this threshold to distinguish substantial exposure to a document's content from a brief mention or summary. It also reflects the DI structure, where each document is preceded by a short descriptive paragraph before the main content. The index document also contains a summary of the document, which therefore allows extracting information without actually visiting it. A \textit{transition} represented movement between source documents. In DI,
transitions are link traversals. In QAI, transitions occurred when the source-document mapping changed within a response or across follow-up responses. Although QAI transitions were mediated by the system's retrieval, they remained user-driven as they resulted from follow-up and compound questions. Self-loops were excluded for comparability, where rereading the currently selected document in DI does not generate a new transition. 

With the visits data, we constructed graphs that describe the navigation pattern of each session. Topics are nodes and transitions are edges. From the graphs, which we plotted for visual qualitative analysis, we additionally extracted the following quantitative measurements (all common graph measurement described in~\cite{weth_dobbs_2013, broder_graph_2000, yoon_comparative_2016, nousiainen_concept_2010}:
\begin{itemize} 
\item\textit{Density:} Measures of overall interconnectedness. Higher values correspond to more tightly connected graphs. 
\item\textit{Directed Diameter:} Measures the breadth of the graph. Lower values correspond to more tightly connected graphs. 
\item\textit{Degree of Variance:} Measures the evenness of connectivity. Higher values correspond to more unevenly connected graphs. 
\item\textit{Average shortest path length:} Measures the average number of steps needed to move between concepts. Lower values correspond to more tightly connected graphs. 
\end{itemize}

\subsubsection{Concept Maps}
In Phase 2, we documented each topic and subtopic mentioned by the participant, and the explicit connections they made to form the concept maps. We collected this data as graphs, which we visualized and analyzed to examine the structure of gained knowledge and mental models following Paul~\cite{paul_analyzing_2014} and Haque et al.~\cite{haque_mental_2023}.

\textit{Topics and Subtopics}: A \textit{topic} corresponds to overarching categories in the worlds represented in a document (\textit{e.g.}, governance, geography (map), or the community structure). \textit{Subtopics} refer to subsections within documents or facts that fit under a topic such as names, roles, dates, or landmarks. We counted a topic/subtopic when the participant made an unambiguous mention that could be mapped to the predefined documents. We did not count repeated mentions of the same topic/subtopic as new instances, only newly stated connections.  References to information beyond the content were marked as inaccurate by the researcher, ensuring completeness since they could still link to new topics or subtopics.

\textit{Number of Connections}: We counted the explicit connections from the topics/subtopics mentioned by participants and any additional connections were verified by the researcher through probing. We also calculated the number of connections per topic.

\textit{Error Metrics}: We counted errors in the elicitation tasks to assess the alignment of participant responses with the actual content of the documents. Errors included statements that contradicted the source documents, unsupported inferences participants made without evidence, and incorrect recall of specific details. Repetitions of the same error were not counted as new, unless they added a new incorrect connection or distinct claim. 

\textit{Graph Metrics}: We calculated the same structural metrics as with the exploration graphs (Subsection~\ref{traversal data}) to assess the properties of participants' mental models. 

\subsubsection{Decision-based Tasks}
We analyzed the decision-based task using the same metrics as the concept mapping task: topics and subtopics count, number of connections, and error metrics. We did not apply graph measures here because the graphs were substantially sparser and some participants were not able to produce responses. We discuss why in the Results section. 


\subsubsection{Qualitative Responses}
The qualitative analysis drew on data sources across research questions. For RQ1, we used participants' interaction logs, queries, QAI responses, and graph visualizations, with relevant interview responses to understand exploration patterns, prompting strategies, and differences between interfaces. 

For RQ2 and RQ3, we conducted thematic analysis~\cite{braun_thematic_2023, braun2021thematic} of the interview reflections and comparison (Phase 4), adopting an inductive-deductive coding approach~\cite{cox_qualitative_2008, Adams_Lunt_Cairns_2008, Dix_2020}. Coding was guided by the RQs while remaining open to emerging concepts. To develop the initial codebook, the first author began by open-coding four transcripts from two participants selected to capture variation in interface use and preference. Initial low-level codes were grouped into preliminary categories to code 12 transcripts more. We discovered new codes, and revisited transcripts to apply them. After this stage, the sample included a balanced set of participants who favored each interface. The research team met regularly to refine and develop themes across the categories of codes. The second round involved collectively collapsing redundancies and adding codes that aligned with the RQs. Specifically, we examined whether responses provided explanations for the quantitative results, such as insights on knowledge construction (RQ2). The third round involved revision and grouping codes into broader themes that reflected patterns in the data. We agreed on a final codebook of 93 codes that was applied deductively to the remaining 16 transcripts, with earlier transcripts revisited as needed to ensure consistency across the data. 

Additionally, participants made direct comparisons by selecting which interface they preferred overall, which they believed supported greater exploration, more learning, and enabled more confident answers. These perceptions were followed by open-ended prompts asking to explain their choices. 

\subsection{Positionality Statement}

None of the authors identify as BLV. Nevertheless, we intentionally ground our study design, analysis, and interpretation in participants' own accounts, practices, and perspectives. Our work centers around the lived realities of SR users, ensuring that RQs, metrics, and conclusions reflect what participants identify as meaningful and relevant. 
The second author has more than six years of sustained engagement with BLV communities. With the fourth author's experience in assistive technology design, these informed the study design, accessibility of the study materials and attention to the screen-reader practices of participants. 
The second author's qualitative research background guided the coding process. 
The third author's expertise in knowledge elicitation and interaction techniques informed our RQs, design and interpretation of elicitation tasks, and the analysis of how participants organized and applied information. The quantitative expertise of the third and fourth authors informed the selection and interpretation of our statistical analyses. 
We view technology as an active mediator that shapes how people access, explore, and make sense of information. Drawing on the relational model of disability~\cite{Thomas01012004, Reindal_2008}, while technology can reduce barriers, we believe restrictive design choices can also actively create disability and impose dysaffordances~\cite{10.7551/mitpress/12255.001.0001}.

%% file: sections/040-Results.tex
\section{Results}
This section presents the study evidence organized by the RQs. Within each subsection, we consider first analysis of quantitative measurements and then evidence from a qualitative analysis.

\subsection{How do People Explore Information Spaces with the Different Interfaces? (RQ1)}

\subsubsection{Quantitative Results.} 
\begin{figure}[t]
  \centering
  \includegraphics[width=\imgw]{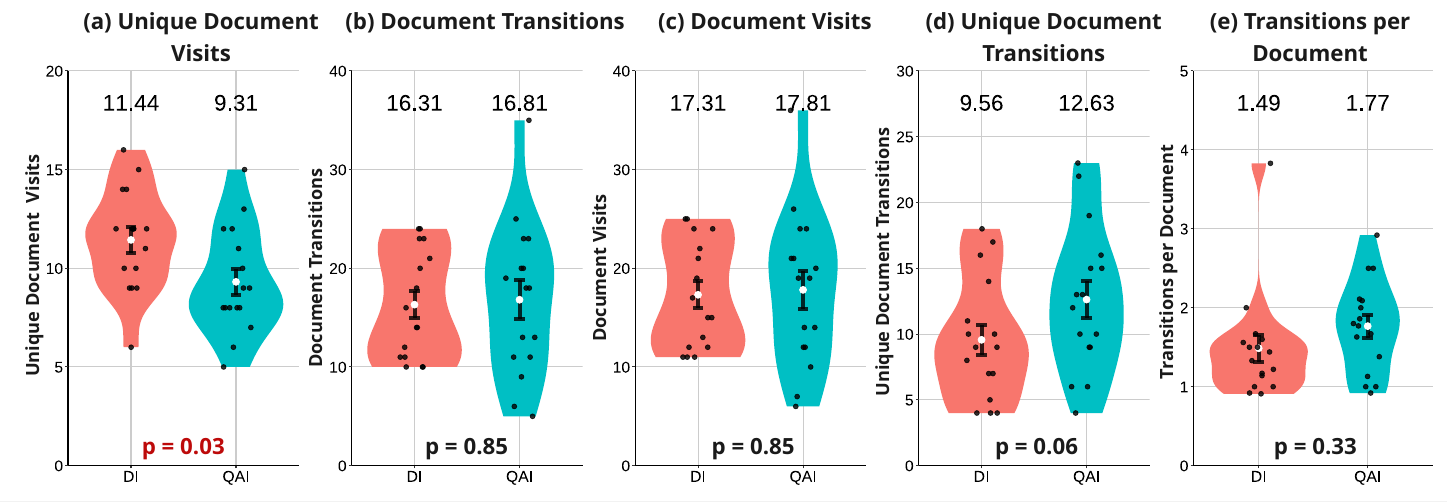}
  \caption{Exploration metrics (RQ1---Phase 1). DI is red, and QAI blue. Numbers above represent the averages. Error bars are standard error. Each black dot is a session.}
  \Description{Violin plots comparing DI and QAI on five measures. Unique document visits averaged 11.44 for DI and 9.31 for QAI, p = 0.03. Document transitions averaged 16.31 for DI and 16.81 for QAI, p = 0.85. Document visits averaged 17.31 for DI and 17.81 for QAI, p = 0.85. Unique document transitions averaged 9.56 for DI and 12.63 for QAI, p = 0.06. Transitions per document averaged 1.49 for DI and 1.77 for QAI, p = 0.33.}
  \label{fig:rq1-exploration-metrics}
\end{figure}

\begin{figure}[t]
  \centering
  \includegraphics[width=\imgw]{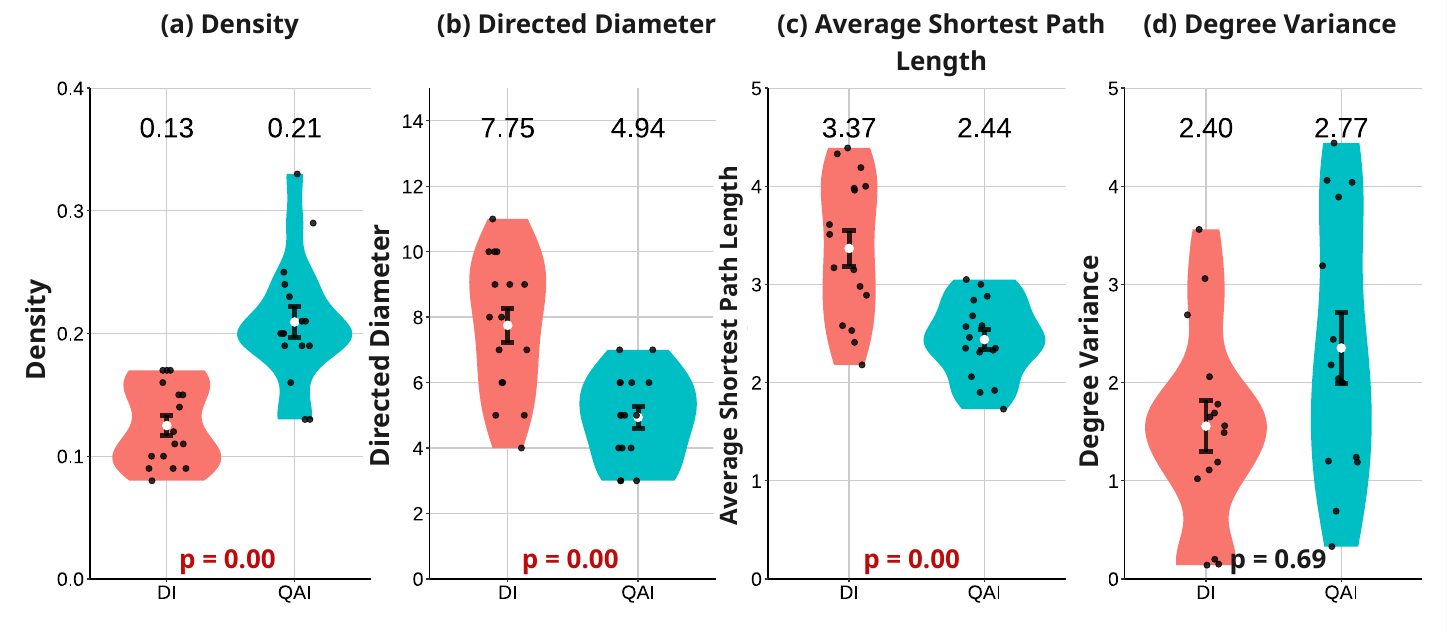} 
  \caption{Exploration graph metrics (RQ1---Phase 1). See caption of Figure~\ref{fig:rq1-exploration-metrics} for details.}
  \Description{Violin plots comparing DI and QAI on four graph measures. Density averaged 0.13 for DI and 0.21 for QAI, p = 0.00. Directed Diameter averaged 7.75 for DI and 4.94 for QAI, p = 0.00. Average Shortest Path Length averaged 3.37 for DI and 2.44 for QAI, p = 0.00. Degree Variance averaged 2.40 for DI and 2.77 for QAI, p = 0.69.}
  \label{fig:rq1-graph-metrics}
\end{figure}

The quantitative evidence for RQ1 is based on statistical comparisons of key measurements represented in Figure~\ref{fig:rq1-exploration-metrics} and Figure~\ref{fig:rq1-graph-metrics}. When counting the unique visited documents, we see that participants tended to visit more documents with the DI ($\mu_{documents,DI}=11.44$, $\sigma_{documents,DI}=2.58$) than with the QAI ($\mu_{documents,QAI}=9.31$, $\sigma_{documents,QAI}=2.68$), a 22.9\% increase, ($t(15)=2.34$, $p=0.03<0.5$,$\eta^2=0.27$). Because documents correspond to topics, this generally supports that exploration was broader with the DI than with the QAI. This is despite the fact that our measurements of total document visits were similar (participants did roughly the same number of visits on average).

We expected that the QAI's nature would lead to more transitions between documents, or at least, more different instances of transitions between documents/topics. This is a property inherent to the QAI: when answering a query, the LLM can output material from multiple documents within the same answer (which we counted as transitions), whereas in the DI transitions can only be done in pre-established ways, such as navigating using hyperlinks between the documents or through the index document.
Despite this, the counts of transitions between documents were not conclusively different (all $p>0.05$ in Figure~\ref{fig:rq1-exploration-metrics} b, d and e). However, a second set of measurements focuses on whether differences in the interface resulted in a tighter interconnection of more distant topics (Figure~\ref{fig:rq1-graph-metrics}). This time the answer is positive. When measuring the density, directed diameter and average shortest path length of the navigation graphs (nodes are topics/documents and links are transitions between those topics) we see that the QAI enabled tighter connections between different topics as measured by a higher density of the navigation graph for QAI ($\mu_{density,QAI}=0.21$, $\mu_{density,DI}=0.13$), a lower directed diameter ($\mu_{diameter,QAI}=4.94$, $\mu_{diameter,DI}=7.75$), and a lower average shortest path length between two topics ($\mu_{aspl,QAI}=2.44$, $\mu_{aspl,DI}=3.77$). 

\subsubsection{Qualitative Results.}
We used answers from the interviews, visualizations of the exploration patterns, and analysis of the participants' questions and the QAI's answers to make sense of how the two interfaces differ and to contextualize the quantitative results from the previous section. We report counts (e.g., $n=5$, where $n$ denotes the number of distinct participants that fit the finding) to make the distribution of coded observations transparent, facilitate comparison across findings, and strengthen analytical credibility~\cite{nicmanis2024reflexive, noble2025ensuring}. We assessed the importance of each theme based on its relevance to our research questions and its relationship to the observed differences.

Our qualitative analysis of participants' explorations revealed three recurring subpatterns: \textit{overview} (learning the overall structure of the information space), \textit{depth} (learning in increasing detail about a particular topic), and \textit{fact finding} (finding answers to specific questions). 
These align with existing frameworks in exploratory search, information foraging, and sensemaking literature~\cite{shneidermanEyesHaveIt1996, marchionini_2006, marchioniniFindingFactsVs1988} where similar knowledge-oriented intents have been documented as fundamental for complex information spaces. While these behaviors represent general patterns of engagement, we observed that for both \textit{overview} and \textit{depth} it did matter whether participants were content with the \emph{predetermined} structure provided in the documents (i.e., the hierarchical organization and sequential order provided by the document author) or wanted to access the information in a \emph{self-defined} organization. For fact finding this does not seem to matter that much.

For example, in our worlds, the documents are structured according to a traditional textbook-like set of topics such as history, politics, or culture. However, there are other ways to slice and index the same information space, such as organizing the descriptions as a storyline, or focusing on the influence of specific historical figures. Although both interfaces can support any of the subpatterns (overview, depth, fact finding), the data suggests that the effectiveness of the interface depends largely on whether the participant is ready to leverage the pre-defined structure of the documents or seeks a different one, as reflected in Table~\ref{tab:rq1-exploration-patterns}. The following paragraphs discuss evidence for each of the Table's cells. 

\begin{table}[]
\resizebox{\columnwidth}{!}{%
\begin{tabular}{|c|c|c|c|c|c|}
\hline
             & \textbf{\begin{tabular}[c]{@{}c@{}}Overview\\ Predetermined\end{tabular}} & \textbf{\begin{tabular}[c]{@{}c@{}}Overview\\ Self-Defined\end{tabular}} & \textbf{\begin{tabular}[c]{@{}c@{}}Depth\\ Predetermined\end{tabular}} & \textbf{\begin{tabular}[c]{@{}c@{}}Depth\\ Self-Defined\end{tabular}} & \textbf{\begin{tabular}[c]{@{}c@{}}Fact\\ Finding\end{tabular}} \\ \hline
\textbf{DI}  & \cellcolor[HTML]{5FF55D}Easy                                              & \cellcolor[HTML]{F78E87}Hard                                             & \cellcolor[HTML]{5FF55D}Easy                                           & \cellcolor[HTML]{F78E87}Hard                                          & \cellcolor[HTML]{F78E87}Hard                                    \\ \hline
\textbf{QAI} & \cellcolor[HTML]{F78E87}Hard                                              & \cellcolor[HTML]{5FF55D}Easy                                             & \cellcolor[HTML]{F78E87}Hard                                           & \cellcolor[HTML]{5FF55D}Easy     & \cellcolor[HTML]{5FF55D}Easy                                    \\ \hline  
\end{tabular}
}
\caption{Summary of the difficulty of using the DI and QAI across the exploration subpatterns, based on qualitative data.}
\Description{Qualitative summary of which interface participants found easier for each exploration intent. The table compares DI and QAI across overview and depth tasks under two conditions: predetermined and self-defined. The table also includes fact finding. Green cells mark the interface reated easier for that subpattern; red cells mark the interface rated harder.}
\label{tab:rq1-exploration-patterns}
\end{table}

\textit{Overview-Predetermined.} Participants ($n=9$) identified advantages of the DI when they were willing to accept the existing structure of the documents. P2 observed that they \textit{``could look at a bunch of categories and decide what to read,''} reflecting the traditional way to familiarize oneself with the information space, akin to skimming the main subsections of a Wikipedia page. P12 echoes this: \textit{``The information was there. It told me the links, told me 25 in a list, gave me an idea of what I was getting into.''} 
In contrast, the QAI presents some difficulty, since a predetermined organization is not immediately apparent, and participants ($n=9$) did not necessarily know how to obtain an overview. P4 observes: \textit{``I didn't know what questions to ask it because I didn't know what information would be available to me.''} Some participants did not even realize that an overview could be useful ($n=6$), as P7 reflects, \textit{``I have no idea what the parameters are. I guess I could have asked it a question like, give me the main topics.''}

\textit{Overview-Self-Defined.} When participants ($n=5$) are, instead, interested in their particular overview of the information, DI can be frustrating because it requires them to define their own categories: \textit{``It felt like reading a book and trying to navigate through the different pages.'' (P11)} This is much easier with the QAI, because participants ($n=7$) can ask the interface to organize information based on their interest. P1 explains that they got \textit{``to determine what I was going to look into and get those big concepts. The other interface [DI] was a bit overwhelming,''}. With QAI, they were able to request cross-document summaries based on their overview preferences without having to check every document.

\textit{Depth-Predetermined.} When participants were looking for depth, opinions are parallel. DI is great if they do not have a particular view on how to dig deeper, and are comfortable following the document's structure ($n=7$). Referring to what a document has to offer, P10 notes: \textit{``I know there's a container full of information in each that I can just explore.''} P12 echoes this, \textit{``at least there was a container that I knew all the information was in, and I knew how big it was.''} 
In contrast, this is harder to do with QAI ($n=8$): \textit{``What happens if I keep asking for more detail, it’s eventually gonna run out of information about this one thing, so now what?'' (P11)} Even when an answer felt detailed, the nature of the question-answer cycle makes it difficult to know when a topic has been exhausted. 

\textit{Depth-Self-Defined.} Nevertheless, if participants are not interested in the grouping (or ``narrative''), then DI required readers to manually trace their path across documents and pursue a topic in detail based on what they deemed relevant ($n=9$). P4 observes, \textit{``...it led me to a whole different path of where I wanted to go to understand the world.''} This made self-defined depth possible, but  more effortful to keep track of. Conversely, the QAI strongly supports presenting information in whichever way the participant asks ($n=7$). For example, participants asked questions such as \textit{``How is life governed according to the full cycle?'' (P4)} or \textit{``Can you explain to me how is it from 12 to 50 years?'' (P9)}. P16 explicitly highlights this advantage, \textit{``you can just keep asking questions and isolate on specific topics,''} without having to look for relevant pieces from each document. 

\textit{Fact Finding.} There is no distinction between whether participants are interested or not in the predetermined organization of the document. The QAI ($n=9$) easily provides contextualized facts as answers (i.e., and advanced search feature) that the DI would require the reader to hunt throughout multiple documents ($n=7$). This advantage with QAI over DI is acknowledged by P4, \textit{``you're able to get your answers as opposed to having to read 30 pages to find out.''}

\subsubsection{RQ1 Synthesis}
The quantitative data show that DI enables a wider coverage of the information space, whereas QAI supports a more tightly connected set of transitions between the different topics. The qualitative analysis helps us connect those results to the fundamental differences between the interfaces: there is value in having access to a pre-established structure that readers can just consume, which is much easier to do with DI. Participants struggle to access structure with QAI because they have to come up with appropriate queries about a structure and information space that they do not know much about in advance. However, when the user wants to access the information space in more sophisticated or custom ways, such as based on personal interest, then the QAI can significantly facilitate this process by crafting coherent answers that comprise information scattered around different areas of the document.

\subsection{How do the Different Interfaces Influence Integration of Knowledge? (RQ2)}
If RQ1's results address how people behaved with the interfaces, RQ2 focuses on the consequences. Were participants able to build, connect, and apply what they explored, integrating it into their mental models? (Phase 2) or when using it in scenario-based responses (Phase 3)? We look at each of the two phases quantitatively first, then discuss the qualitative evidence together.

\subsubsection{Concept Maps (Phase 2)}
\begin{figure}[t]
  \centering
  \includegraphics[width=\imgw]{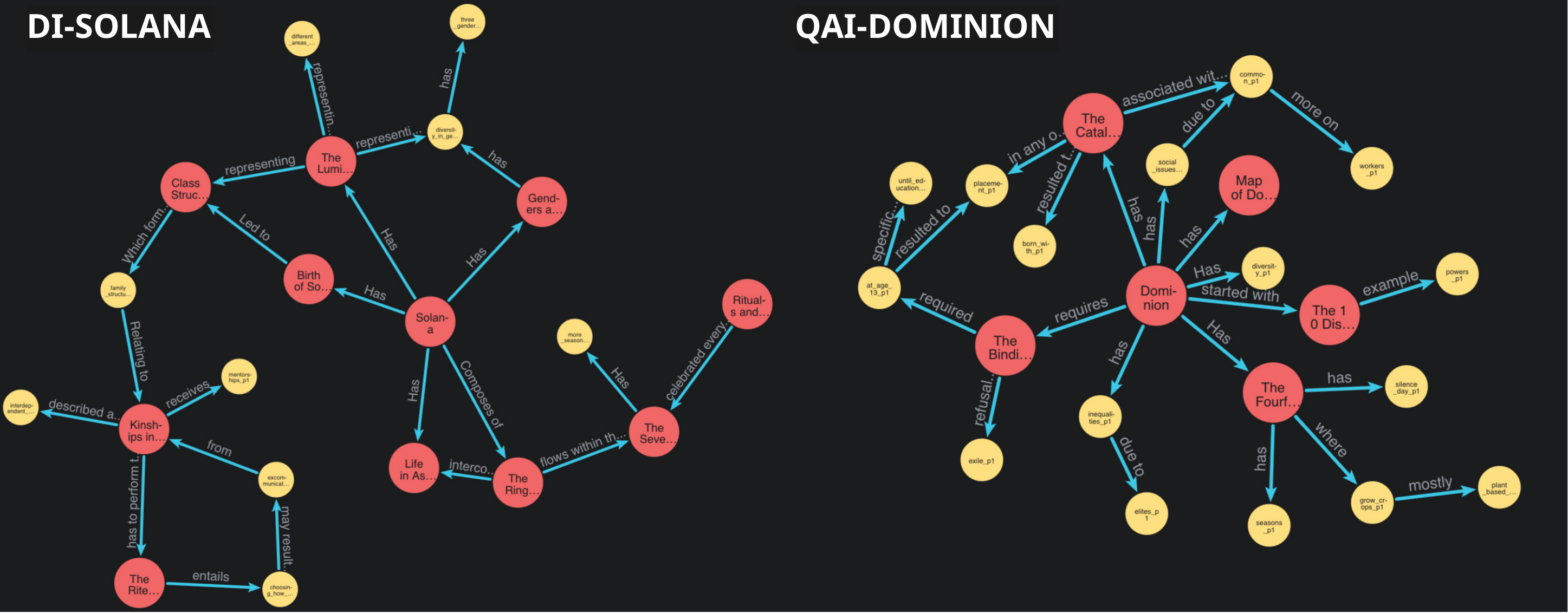} 
  \caption{Participant 1's mental model graphs from the concept mapping. DI (left) and QAI (right). Nodes represent elicited topics (red) and subtopics (yellow). Edges represent explicitly described relationships.}
  \Description{On the left, the DI-Solana graph shows interconnected nodes arranged around central topics. There are 11 topics and 9 subtopics. On the right, the QAI-Dominion graph shows 6 topics and 16 subtopics with a more centralized structure with hub nodes connecting to many others. Nodes represent topics or subtopics, and arrows represent directed relationships between them.}
  \label{fig:MMGraphs}
\end{figure}

\begin{figure}[t]
  \centering
  \includegraphics[width=\imgw]{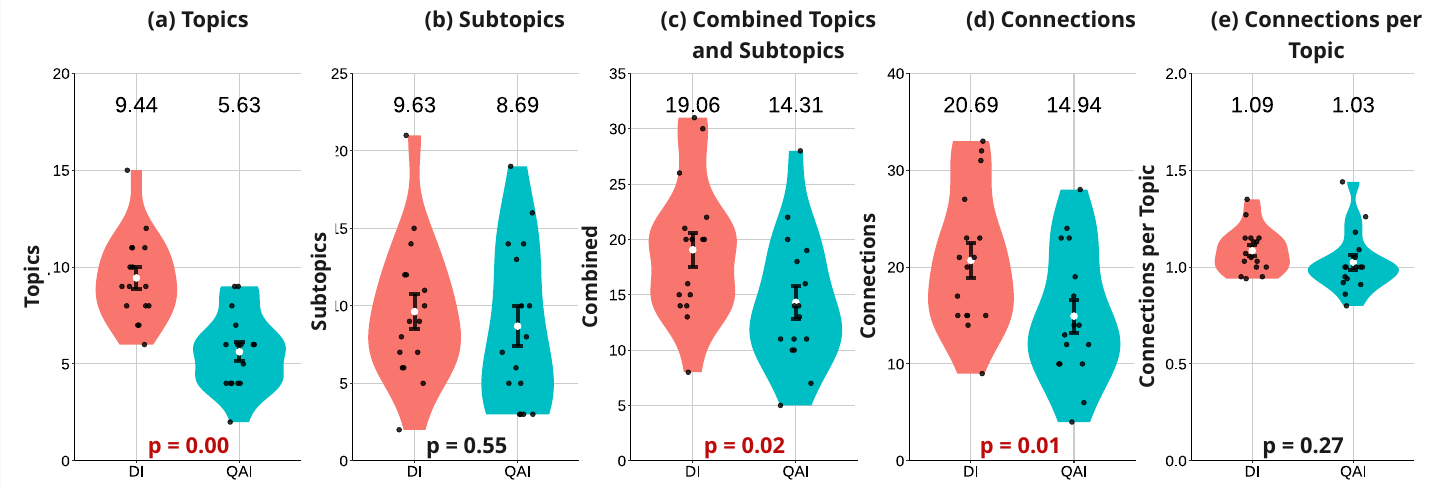} 
  \caption{Concept map metrics (RQ2---Phase 2). See caption of Figure~\ref{fig:rq1-exploration-metrics} for details.}
  \Description{Violin plots showing five metrics for DI and QAI. Topics averaged 9.44 for DI and 5.63 for QAI, p = 0.00. Subtopics averaged 9.63 for DI and 8.69 for QAI, p = 0.55. Combined topics and subtopics averaged 19.06 for DI and 14.31 for QAI, p = 0.02. Connections averaged 20.69 for DI and 14.94 for QAI, p = 0.01. Connections per topic averaged 1.09 for DI and 1.03 for QAI, p = 0.27}
  \label{fig:concept-maps-data}
\end{figure}

\begin{figure}[t]
  \centering
  \includegraphics[width=\imgw]{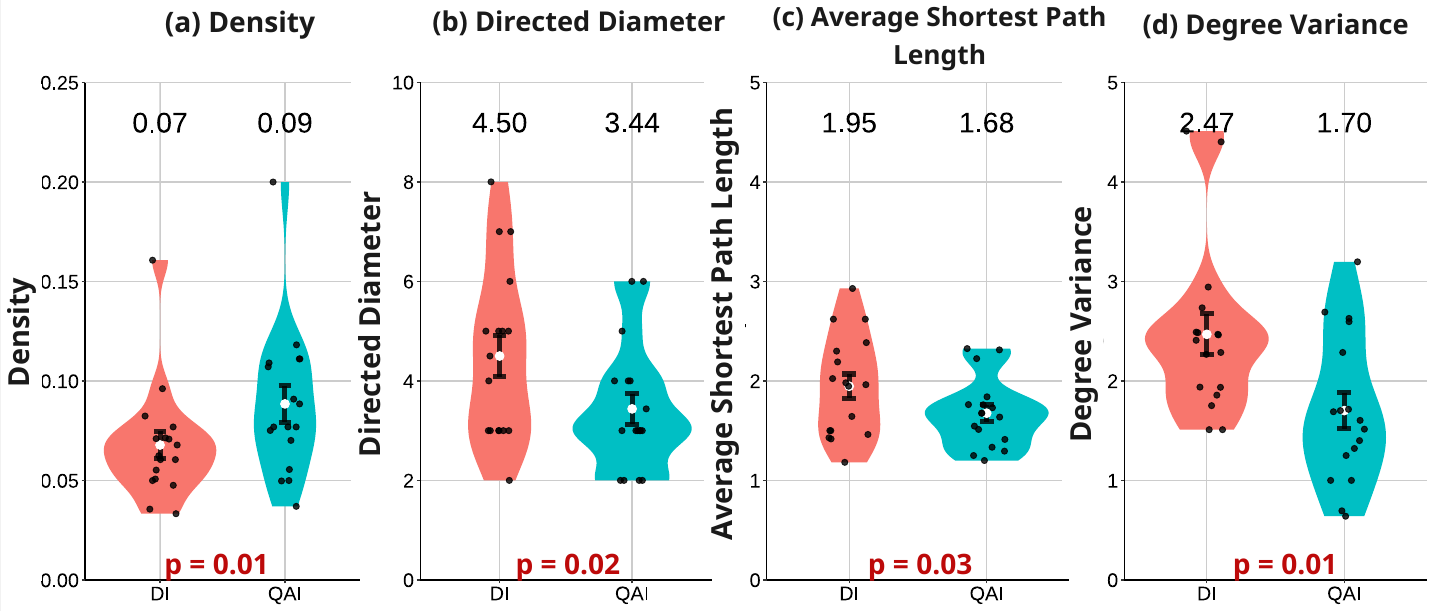} 
  \caption{Concept map graph metrics (RQ2---Phase 2). See caption of Figure~\ref{fig:rq1-exploration-metrics} for details.} 
  \Description{Violin plots showing four metrics for DI and QAI. Density averaged 0.07 for DI and 0.09 for QAI, p = 0.01. Directed diameter averaged 4.50 for DI and 3.44 for QAI, p = 0.02. Average shortest path length averaged 1.95 for DI and 1.68 for QAI, p = 0.03. Degree variance averaged 2.47 for DI and 1.70 for QAI, p = 0.01.}
  \label{fig:concept-maps-graphs}
\end{figure}

We asked participants to recreate their mental structures of the worlds they had explored (the elicitation process is described in Section~\ref{study procedure}). The result of this process is a mental model graph for each session and participant that reflects their understanding of the information space (Figure~\ref{fig:MMGraphs} displays two examples). The graphs are analyzed in three ways: straight counts of topics, subtopics, and their links (Figure~\ref{fig:concept-maps-data}), metrics of their connectedness (Figure~\ref{fig:concept-maps-graphs}), and the correctness of the remembered topics and links (Figure~\ref{fig:error-metrics}). 

Overall, participants included 67\% more topics on average in their elicited mental model graphs in the DI condition ($\mu_{topics,DI}=9.44$) compared to the QAI condition ($\mu_{topics,QAI}=5.63$, $t(7.25)$, $p=0.02$, $\eta^2=0.78$) and 11\% more subtopics ($\mu_{subtopics,DI}=9.63$, \\ $\mu_{subtopics,QAI}=8.69$, $t(0.61)$, $p=0.55$, $\eta^2=0.02$). Considering topics and subtopics together, this represents evidence that the models elicited in the DI condition were more comprehensive on average. With DI, participants made 38\% more connections ($\mu_{connections,DI}=20.69$, $\mu_{connections,QAI}=14.94$, $t(2.92)$, $p=0.01$, $\eta^2=0.36$), but this seems to be attributable to having more nodes to connect (the connections per topic measures are fairly similar and not statistically distinguishable: $\mu_{connections/topic,DI}=1.09$, $\mu_{connections/topic,QAI}=1.03$, $t(1.15)$, $p=0.27$, $\eta^2=0.08$).

We also considered each node or link and classified it as correct or incorrect with respect to the knowledge contained in the documents. In terms of raw error counts, participants made 45\% fewer errors with DI than with QAI ($\mu_{errors,DI}=3.31$, $\mu_{errors,QAI}=6.00$, $t(5)$, $p<0.01$, $\eta^2=0.62$). Because the elicited models differed in size, we used error rate to show that the difference is slightly amplified (54\%) ($\mu_{error rate,DI}=0.12$, $\mu_{error rate,QAI}=0.26$, $t(6.10)$, $p<0.01$, $\eta^2=0.71$).

We then calculated graph metrics for the concept maps, after removing errors. The results show that the DI elicited mental model graphs were less interconnected than QAI's, with less average density, higher diameter, higher average shortest path length and higher degree variance (all $p<0.04$, see Figure~\ref{fig:concept-maps-graphs}).

\subsubsection{Decision-based Scenario (Phase 3)}
\begin{figure}[t]
  \centering
  \includegraphics[width=\imgw]{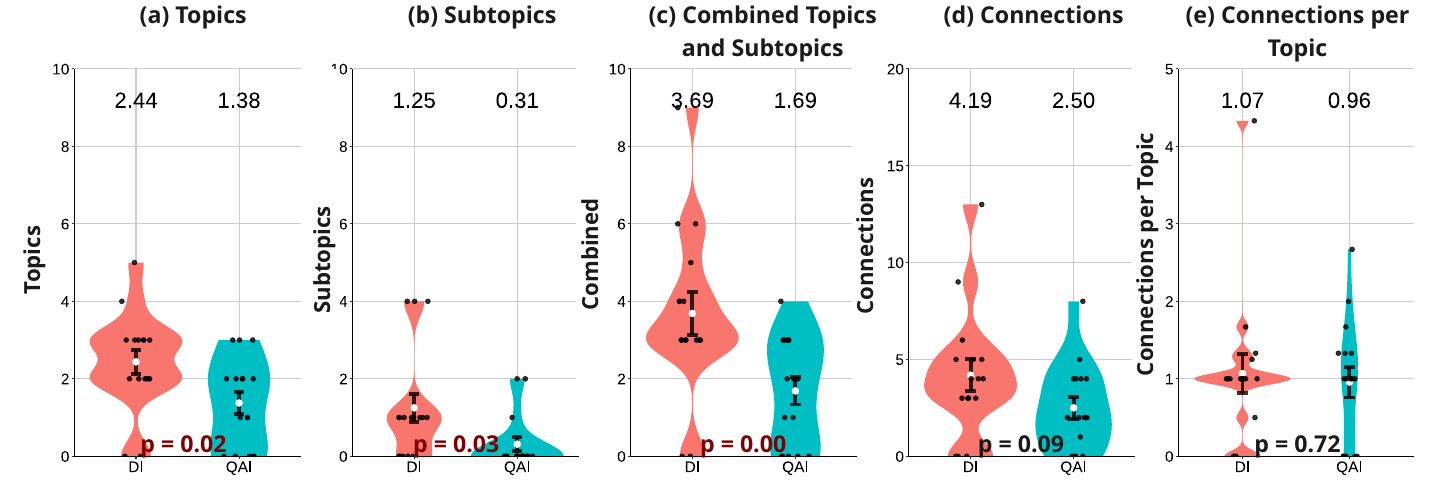} 
  \caption{Decision-based metrics (RQ2---Phase 3). See caption of Figure~\ref{fig:rq1-exploration-metrics} for details.}
  \Description{Violin plots showing five metrics for DI and QAI. Topics averaged 2.44 for DI and 1.38 for QAI, p = 0.02. Subtopics averaged 1.25 for DI and 0.31 for QAI, p = 0.03. Combined topics and subtopics averaged 3.69 for DI and 1.69 for QAI, p = 0.00. Connections averaged 4.19 for DI and 2.50 for QAI, p = 0.09. Connections per topic averaged 1.07 for DI and 0.96 for QAI, p = 0.72}
  \label{fig:decision-based-data}
\end{figure}

\begin{figure}[t]
  \centering
  \includegraphics[width=\imgw]{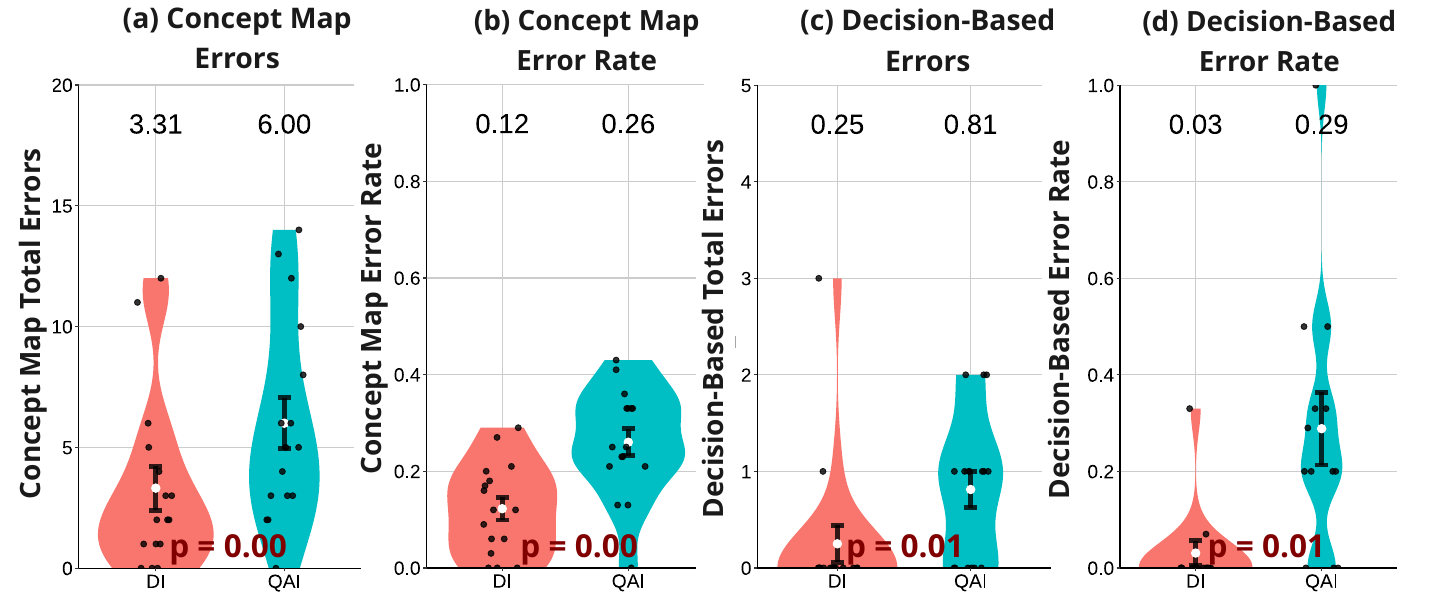} 
  \caption{Concept map and Decision-based error metrics (RQ2---Phase 3). See caption of Figure~\ref{fig:rq1-exploration-metrics} for details.}
  \Description{Violin plots comparing DI and QAI across four error metrics. Concept map errors averaged 3.31 for DI and 6.00 for QAI (p = 0.00), while concept map error rates averaged 0.12 for DI and 0.26 for QAI (p = 0.00). Decision-based errors averaged 0.25 for DI and 0.81 for QAI (p = 0.01), and decision-based error rates averaged 0.03 for DI and 0.29 for QAI (p = 0.01).}
  \label{fig:error-metrics}
\end{figure}

We performed a parallel analysis of the participant answers to the scenario phase, where we asked participants to apply their knowledge of the worlds to suggest solutions to conundrums. The results follow very similar patterns as the previous subsection, which instead of describing in text we leave summarized in Figures~\ref{fig:decision-based-data} and~\ref{fig:error-metrics}. However, there are a few differences between the two analyses. First, the scenario task was much harder than the elicitation task, which left 6 participants unable to provide coherent responses in one ($n=5$, 4 of which were with the QAI) or both ($n=1$) of the two sessions. Second, the pattern of higher connections in DI was not strong enough for the conditions to be distinguished statistically. We did not run the graph metrics for the resulting graphs because these graphs are much sparser than in Phase 2, hence noisier and less meaningful. 

\subsubsection{Qualitative Results}
We used interview responses to analyze how participants made sense of the worlds using the two interfaces. Our qualitative analysis involved three processes: \textit{building} understanding of the information space, \textit{connecting} information across documents, and \textit{applying} their gained knowledge to a scenario-based question. 

\textit{Build.} Participants described clear differences in how the two interfaces supported building an understanding of the created worlds. Participants appreciated that DI did not require them to construct their own structure since it is already visible ($n=5$). P3 explains that DI \textit{``gives you an understanding of what each chapter is and it shows you what you're looking, what you're headed into.''} However, DI can be frustrating if readers want to build their own mental structure ($n=4$). P10 notes, \textit{``I basically have to figure out the structure based on the description of each thing,''} suggesting that they still had to determine where a given piece of information fit in their mental models. Conversely, the QAI supported participants in \textit{``being able to curate the index in my own thought'' (P3)} by building structure through their questions ($n=7$). At the same time, this flexibility depended on participants having skills on what and how to ask ($n=10$), \textit{``it was a bit more tricky because I had to have a bit more imagination.''(P6)}

\textit{Connect.} When it comes to integrating disparate pieces of information across the information space, some participants perceived the DI as providing more information volume which, in turn, allowed them to establish more connections between the different parts of the information space ($n=6$). For example P3 explains that, with DI, \textit{``I felt like I had more to work with and it was more that I could kind of jump off of, so it was helpful.''} Yet for some, most connections with DI are implicit (i.e., not links), which forces them to make the connections mentally ($n=3$). Even when the connection is explicit (i.e., a link), it might still be difficult to know why they are connected. P7 explains this: \textit{``if you're going through it as link to here, link to here, you're not necessarily getting the connections between those.''}.

In contrast, several participants recognized that the QAI directly supports integration of related topics, unconstrained by the structure ($n=7$). In other words, participants appreciate that the QAI composes a coherent text using different bits of information: \textit{``AI does a better job of linking stuff together than you having to pick a piece and go to another link. It pulls it together.'' (P7)} However, this is not for free, since participants recognized that asking the right questions to get the right integration is not trivial ($n=8$): \textit{``my questions were probably pretty simplistic and it sort of led you around in circles a couple of times. It's just like any AI''} (P8).



\textit{Apply.} The decision-based scenario (Phase 3) interrogated whether participants could use the knowledge they had gained and connected through each interface. As we mentioned before, this is likely a much harder task that goes beyond accessing and remembering and requires integration, retrieval, and even creativity; in other words, it is a task higher in Bloom's taxonomy~\cite{krathwohl_revision_2002}. Some participants thought that the stable and well-organized structure of documents (DI) helped them answer the scenario ($n=8$). P4 noted about the information to answer a question that \textit{``it gave more, helped me get a visual representation first of what was happening with the various structures.''}
However, some participants also recognized that, to be successfully applied, they needed to actively process the information: \textit{``It was much more of a lecture because you were just absorbing; I was filling the gaps,''} (P11) ($n=3$).

With QAI the challenge is of a different nature. Although participants thought that they could ask for information that meaningfully supported their progressive understanding and ability to apply the knowledge ($n=6$---``I was asking questions that were important to me.'' (P15)), they were never sure that what they had gathered through questions and answers was sufficient ($n=8$): \textit{``knowing what to ask and then questioning the output. Am I getting the full answer? Am I getting the correct information?''(P14)}.


\subsubsection{RQ2 Synthesis}
The quantitative data show that participant's elicited mental models contained, on average, more topics and subtopics when using the DI, although less densely connected. Importantly, the elicited graphs also had many more errors. When asked to apply what they had learn, we also observed an advantage when using the DI, which resulted in substantially more topics and subtopics brought up, and also fewer errors. The qualitative data suggests that building mental models with the DI requires effort to connect what they read with their current understanding as they go, discovering implicit connections between the topics. In contrast, the QAI offers more flexibility to build the model based on the participant's own curiosity, and interest. The QAI answers provide custom-formulated answers that directly answer the participant's questions and connect the topics smoothly in the answer's text. In exchange, formulating the right questions might be challenging, and the participants were often not sure of whether they had gathered enough information or reached exhaustive cover of the available information.

\subsection{What Aspects of the Different Interfaces Affect Preferences? (RQ3)}
At the end of Phase 4, we asked participants for comparative reflections on their experience with the interfaces. Table~\ref{tab:perception-measurent-match} summarizes these interview responses together with counts of their actual performance, and the counts of alignment. Overall, participants' perceptions did not fully align with the measured patterns. This discrepancy was evident for breadth of exploration and appeared even more strongly for perceived support for learning and confidence in answering the questions. When participants chose their preferred interface, 8 participants favored each. There is no obvious mapping between perceived advantages on either exploration, mental model building or scenario answer confidence and the final stated preference.

\begin{table}[]
\resizebox{\columnwidth}{!}{%
\begin{tabular}{|c|cc|ccccc|}
\hline
               & \multicolumn{2}{c|}{\textbf{Perception}} & \multicolumn{3}{c|}{\textbf{Measurement}}                                             & \multicolumn{2}{c|}{\textbf{Match}}  \\ \hline
               & \multicolumn{1}{c|}{DI}  & QAI  & \multicolumn{1}{c|}{DI} & \multicolumn{1}{c|}{QAI} & \multicolumn{1}{c|}{Equal} & \multicolumn{1}{c|}{DI} & QAI \\ \hline
\textbf{Explore more}      & \multicolumn{1}{c|}{10}  & 6    & \multicolumn{1}{c|}{11} & \multicolumn{1}{c|}{4}   & \multicolumn{1}{c|}{1}     & \multicolumn{1}{c|}{6}  & 2   \\ \hline
\textbf{Learn more}       & \multicolumn{1}{c|}{6}   & 10   & \multicolumn{1}{c|}{11} & \multicolumn{1}{c|}{5}   & \multicolumn{1}{c|}{0}     & \multicolumn{1}{c|}{5}  & 4   \\ \hline
\textbf{Answer Confidence}  & \multicolumn{1}{c|}{7}   & 9    & \multicolumn{1}{c|}{12} & \multicolumn{1}{c|}{1}   & \multicolumn{1}{c|}{3}     & \multicolumn{1}{c|}{5}  & 0   \\ \hline
\textbf{Overall Preference}& \multicolumn{1}{c|}{7} & 8 & \multicolumn{1}{c|}{N/A} & \multicolumn{1}{c|}{N/A} & \multicolumn{1}{c|}{N/A} & \multicolumn{1}{c|}{N/A} & N/A \\ \hline
\end{tabular}
}
\Description{Comparison of participant judgments and measured outcomes for DI and QAI. Perception shows reported advantages, Measurement shows metric-based advantages, and Match shows agreement between the two. Overall preference is subjective only.}
\caption{Comparison of subjective judgments and measured outcomes for DI and QAI. Perception shows how many participants said each interface helped them explore more (RQ1), learn more and answer with greater confidence (RQ2). Measurement shows how many participants actually performed better with each interface. 
Overall preference is subjective only.}
\label{tab:perception-measurent-match}
\end{table}

Nevertheless, two main themes came up in their interviews and feedback that can help us understand participant preferences: Agency/Control and Effort/Cognitive Load. Interestingly, arguments regarding each of those topics were used to support preferences for either of the two interfaces.

\textit{Agency and Control.} Participants often evaluated the interfaces in terms of how independent they felt in steering the interaction and shaping control around their own interests. With DI, participants valued having the option to decide for themselves what documents to open and which links to pursue ($n=9$) \textit{``I felt really empowered. There’s all these concepts and I’m going to go with what speaks to me,'' (P1)}. P10 similarly describes \textit{``I could just jump into it right away. I could be independent with it.''} Still, for some participants, having the freedom to choose has drawbacks ($n=3$), \textit{``I lost control because of opening documents. I had to go back somewhere totally unrelated in my opinion.'' (P4)} In QAI, control meant having the ability to ask anything, which felt open-ended and self-directed ($n=9$). P9 describes this, \textit{``I liked that I could ask the questions and gave me answers. It acknowledged me.''} Similarly, P14 mentions: \textit{``I have more control over what I want to focus on,''} emphasizing the value of the question-answer cycle. However, control also depended on knowing how to steer the conversation productively ($n=6$), as P12 reflects: \textit{``I don’t think I’m asking the right questions to get the information that I wanted.''} 

\textit{Effort and Cognitive Load.} 
Participants often mentioned the differences in effort involved in obtaining, tracking and working with the information with each interface.
Some participants appreciated DI because, as P6 explains, \textit{``there wasn’t a lot of guesswork,''} and they could \textit{``jump into it right away,''} ($n=5$). P13 echoes this by describing DI as \textit{``a straightforward situation,''} and P10 notes that the structure is \textit{``pretty easy to understand.''} However, the effort required with DI was often placed in repeated browsing and tracking of the documents ($n=4$). P7 particularly dislikes \textit{``I hate things that go on and on and send you off to other links. It’s too much,''} with P11 also reflecting on the amount of material, \textit{``I find this very exhausting and like a lot of information.''} Some participants appreciated the QAI's ability to address several topics at once ($n=9$), as P4 observes: \textit{``I was able to ask three questions at once,''} and with P14 describing their experience as \textit{``better, quicker, and clearer.''} Yet the cost of this interaction relies heavily on the reader to probe the information effectively ($n=5$), with P8 describing: \textit{``it could be just insecurity. Trying to figure out the questions and do it in an intelligent way.''} P15 also reflected that the interaction with QAI depends on the user, \textit{``I think the limitation is your mind, right?''}. 

%% file: sections/050-Discussion.tex
\section{Discussion}
In this section we interpret the results above in light of the three research questions stated in the introduction, then connect the results to the emerging knowledge of GenAI-based interfaces for knowledge access, including accessibility. We finish by synthesizing recommendations for practitioners and users and qualifying the results based on the limitations of our study design choices.

\subsection{Answers to RQ's and their connections}
Our behavioral measures of how people chose to explore the information space exposed a trade-off between the two interface styles. Participants used the more traditional DI in a way that resulted in more unique documents visited than the QAI. However, the QAI enabled more flexible transitions from different parts of the information space (RQ1). These differences are not difficult to trace to the characteristics of the interface: the QAI---backed by a current LLM---is able to synthesize answers that connect disparate pieces of information into a smooth narrative, a hitherto difficult to achieve way to navigate information that many people liked and, importantly, allows the reader to personalize their knowledge acquisition.

Unfortunately, the QAI way of interacting with the information space did not seem to translate into more information remembered, better mental models, or an improved ability to apply the gained knowledge or models. In fact, the evidence that we collected supports the opposite: elicited mental models were smaller and contained more errors, although perhaps were a bit more tightly connected (RQ2). This pattern propagated to the application of their understanding, which was also poorer for QAI.
We do not rely only on errors as the main or only measure of our analysis, in particular because task instructions de-emphasized recall. Nevertheless, the error measurements very clearly confirm results from the other measurements that models were overall poorer when using the QAI. Task instructions that explicitly emphasized memorization might have produced different outcomes, and this warrants further study.

We anticipate two plausible causal explanations for the connection between the behavior in access (RQ1) and our appraisals of the gained information and understanding (RQ2). First, the DI requires more effort (some participants brought this up), hence more elaboration which, in turn, results in deeper processing, better memory and more functional mental models~\cite{lockhart_levels_1990}. Second, the QAI relinquishes the predetermined structure in the documents structure (a kind of ready-to-learn model curated by an expert in advance), forcing the reader to consider which mental structure to build and how to build it, then formulate the appropriate questions to fulfill this purpose. This might simply be too difficult and detract from the mental resources required to do the actual learning. This is, essentially, a Cognitive Load Theory argument about germane cognitive load adding to intrinsic cognitive load to overwhelm cognitive capacity~\cite{sweller_cognitive_1988,paas_instructional_1994}. 

Interestingly, the challenges introduced by the QAI did not appear obvious to participants, who often rated it as best for the outcomes where we know they did worse (i.e., they were often wrong---see Table~\ref{tab:perception-measurent-match}). In turn, their preferences were evenly split between the two interfaces (RQ3). This suggests that the nature of the interfaces somehow obscures their value to their users (a failure of metacognition, see the next subsection) or that preference was informed by other factors other than their ability to form more complete mental models and apply them. In any case, it is noteworthy and somewhat alarming that so many participants thought they were better with the interface that actually helped them least. 

\subsection{What does this mean for GenAI interfaces and non-visual information access?}
Our findings extend work on traditional and GenAI search. Recent studies suggest that GenAI-based tools can support exploratory search, better user experience, and lower perceived effort~\cite{liu_conversational_2021,kaushik_comparing_2025,selker_generative_2024,zerhoudi_searchlab_2025,kim_conversations_2025,yang_searchchat_2025, kim_exploratory_2026}. Comparative work between chatbots and search tools afford different kinds of access, and that the strengths depend on the task, domain, and design~\cite{pasquarelli_ai_2025, melumad_experimental_2025}. This is consistent with our study where the QAI was particularly effective for contextualized fact-finding across multiple documents, whereas the DI better supported broad exploration of the corpus. 
However, most prior evaluations focus primarily on search tasks and do not measure how people build mental models. Our results therefore contribute empirical evidence suggesting that the GenAI-enabled conversational interfaces that seem poised to replace traditional search as well as much direct document access, could be detrimental for tasks that are more open, exploratory or contributing to people's expertise and general knowledge. This is particularly important for the BLV community for two reasons. First, GenAI interfaces are being quickly adopted by the community for many tasks~\cite{adnin_i_2024,tang_everyday_2025}, and traditional document access is often particularly tedious and time-consuming with non-visual input-output~\cite{lee_repurposing_2020}. Both make it more likely that they will be used to replace document access. We also suspect that there might be important differences between BLV users and sighted users when using DI vs.\ QAI interfaces, but this will require further research. 

Our results are also consistent with concerns about the metacognitive consequences of LLM use~\cite{tankelevitch_metacognitive_2024}. Experimental work has shown that despite how LLM-based search feels easier and faster, it can also yield superficial learning since users do less of synthesis~\cite{melumad_experimental_2025, tankelevitch_metacognitive_2024, stadler_cognitive_2024, yong2026characterization} 
This is consistent with our findings. First, we did observe that participants experienced the QAI as efficient, flexible, and responsive, yet these benefits did not translate into larger concept maps, or stronger performance in applying what they had learned. Second, participants were often poor at judging which interface actually supported their exploration or learning better. 
Moreover, we observed that the interfaces function not only as a tool for consuming information but also as a mechanism to help users determine what they have covered, what remains unseen, and how complete or reliable their current understanding is. In our study, the DI relies on users to build an internal representation of document architecture through system interaction at the interface/surface level of the stratified model of information access \cite{saracevic1997}. Users must grasp what documents exist and infer what information they contain,
encouraging users to explore documents across various categories to construct this mental model. 
In contrast, QAI encourages users to focus on formulating their exploratory intent as prompts at the affective layer. 
Thus, the user's prompt and early document choices determine the subsequent course of exploration, making coverage and completeness harder to assess. 

\subsection{Implications for users and future directions for practitioners}
Our findings have implications for the adoption of QAI technologies, for the design of future interactive systems and for future research. 

First, we believe that users need to be aware that, despite their own initial appraisals of the value of LLM-based QA technologies for accessing documents and other information spaces, they incur the risk of processing information more superficially and being less able to apply it. This is particularly important for BLV individuals, who often seek technological solutions to avoid being perceived as less effective than their sighted peers~\cite{zhao_accessibility-driven_2026, arslantas_digital_2022, navas-bonilla_inclusive_2025, zhao2026if}.Further confirmation should encourage wider dissemination of this issue, perhaps in educational contexts. These warnings should be qualified because readers are not always looking to form better models or remember more.

Second, some of the negative effects that we observed might be ameliorated or solved by better design. For example, when people are at a loss of where to start, or how to start structuring their own learning, the system might be able to, prompted or unprompted, offer help on how to carry out the process more efficiently. We did not observe this kind of behavior (asking for help about how to explore, learn or interpret), but it is not rare in current LLMs to offer different levels of hints to the user before or after prompting. Future document interfaces should also be able to incorporate notions of desired agency and effort that adapt to user preferences. 

Third, designing new hybrid modes of access for document and information spaces might offer the advantages of DIs and QAIs without the disadvantages, depending on the task or the purpose of the interaction. Our study intentionally separated these modes of access to make differences easier to identify and attribute. Future systems, however, could combine document navigation, search, and conversational interaction to support different tasks and preferences. Interestingly, it might be of research interest to also consider whether documents and other current ways to structure information for human consumption can benefit from different forms and structures, especially if we accept that question-answer interfaces might become the dominant way to access them by the BLV community.

\subsection{Limitations}
We strove to design the QAI and DI interfaces in a way representative of current interfaces and as close to the state of the art as possible. Nevertheless, many different choices to the ones we made are justifiable, and we cannot discard that they could have influenced the results in different ways. One such choice was enabling touch interaction for spatial documents. This means that we cannot separate the effects of multimodality from other aspects of the DI interface. 
In other words, having different types of documents might have improved people's ability to remember and integrate the information. 
Similarly, we decided to include a document in each corpora that provides access to all other documents. This somewhat reduced the incentive for people to navigate between topics from other in-document links, but we also think it is more realistic. Although we still stand by our experimental design choices, we need further experiments to know whether either of these caused the improved performance in the DI condition. Specifically, future systems might include features of both DI and QAI, including both tactile spatial access, search, and presentation of literal information from the document corpus. Ecologically faithful evaluations of such system will require new experiments.

Additionally, we had to decide how to count visits and transitions between visits in Phase 1 in a way that would allow comparisons between the two interfaces. We chose before analysis and with our best judgment. However, we acknowledge that other choices were possible that might result in different balances. Nevertheless, this only affects some measurements in Phase 1; other measurements in other phases are less open to variation because the model elicitation and scenario processes are identical for the two conditions. 

Our findings were not independently reviewed by the BLV community. We also focused on SR users to maintain 
a consistent access modality, as SR use imposes navigation demands central to how these interfaces are experienced, limiting generalizability to BLV users who use other access methods.
Finally, we only tested participants in two sessions of ninety minutes. Prolonged use of interfaces for information access is likely to evolve, both in use patterns and in its outcomes.



%% file: sections/060-Conclusion.tex
\section{Conclusion}
Our study compared two interfaces, DI and QAI, to understand how BLV users explore and build knowledge of unfamiliar domains. From this comparison, we draw attention to a meaningful distinction between interfaces. The nature of interaction with QAI through conversation can feel deceptively expansive, offering a sense of coverage even though content is synthesized. This, in turn, did not translate into better learning. In contrast, DI may demand more effort or might be overwhelming, but they result in a more comprehensive mental model due to the provided structure. This distinction is important because participants do not always recognize it. Many perceived the QAI as better even when performance suggested otherwise. For BLV users, who may be motivated to adopt these tools to reduce the burden of traditional document access, this introduces a risk of interfaces tha feel easier may also support weaker understanding. As LLM-based interfaces evolve closer to becoming intermediaries, accessible technologies should be evaluated beyond speed or convenience, but also accurate understanding and meaningful exploration.